\documentclass[superscriptaddress,a4paper,tightenlines,nofootinbib,floatfix,11pt]{article}

\usepackage{jheppub}

\usepackage{hyperref}
\usepackage{url}
\usepackage{graphics}
\usepackage{amsmath}
\usepackage{cancel}
\usepackage{bm}
\usepackage{graphicx}
\usepackage{subcaption}
\usepackage{booktabs}
\usepackage{siunitx}
\usepackage{amssymb}
\usepackage{relsize}
\usepackage{physics}
\usepackage{chemarrow}
\usepackage{multirow}
\usepackage{soul}
\usepackage[normalem]{ulem}
\usepackage{comment}

\newcommand{\chonepm}{\tilde{\chi}_1^{\pm}}

\def \st1{{\widetilde{t_1}}}
\def \mst1{m_{\widetilde{t_1}}}

\def \lspone{\tilde\chi_1^0}
\def \mlspone{m_{\lspone}}
\def \lsptwo{\tilde\chi_2^0}

\def\chonepm{\tilde{\chi}_1^{\pm}}

\def\mchonepm{m_{\chonepm}}
\def\champ2{\tilde{\chi}_2^{\mp}}

\def \met{\rm E{\!\!\!/}_T}

\def\tcb{\textcolor{blue}}
\def\tcm{\textcolor{magenta}}

\def \ifb{\rm {fb}^{-1}}

\title{Revisiting the Electroweakino Sector of the Baryon Number Violating MSSM at the HL-LHC with Deep Neural Networks}

\author[1]{Rahool Kumar Barman,}
\author[2]{Arghya Choudhury}
\author[2]{Subhadeep Sarkar}
\affiliation[1]{Kavli IPMU (WPI), UTIAS, The University of Tokyo, Kashiwa, Chiba 277-8583, Japan}
\affiliation[2]{Department of Physics, Indian Institute of Technology Patna, Bihar - 801106, India}

\emailAdd{rahool.barman@ipmu.jp}
\emailAdd{arghya@iitp.ac.in}
\emailAdd{subhadeep\_1921ph21@iitp.ac.in}

\abstract{We study the projected sensitivity of direct electroweakino production $pp \to \chonepm \lsptwo$ at the High Luminosity LHC~(HL-LHC: $\sqrt{s}=14~$TeV, $\mathcal{L}=3~\textrm{ab}^{-1}$) in a simplified framework with wino-like, mass degenerate $\chonepm$ and $\lsptwo$, and a bino-like lightest neutralino $\lspone$, assuming R-parity violating~(RPV) through the baryon number violating $\lambda^{\prime \prime}_{112}u^c d^c d^c$ and $\lambda^{\prime \prime}_{113}u^c d^c b^c$ operators. We consider three channels with the $\lambda^{\prime \prime}_{112}u^c d^c d^c$ RPV operator: $Wh$ mediated $1\,\ell + 2\,b + \met$, $Wh$ mediated $1\,\ell + (\geq 2\,j) + 2\, \gamma + \met$, and $WZ$ mediated $3\ell + (\geq 2 j) + \met$. Similar final states also arise from the $\lambda^{\prime \prime}_{212}$ coupling.
We also analyze two channels with the $\lambda^{\prime \prime}_{113}u^c d^c b^c$ RPV operator: $Wh$ mediated  $1\,\ell + (\geq 1\,b\,) + (\geq 1\,j) + 2\, \gamma + \met$, and $WZ$ mediated $3\ell + (\geq 1 b) + \met$. The $\lambda^{\prime \prime}_{123}$, $\lambda^{\prime \prime}_{213}$, and $\lambda^{\prime \prime}_{223}$ couplings also lead to these two final states.
In each channel, we train benchmark-specific multi-layer perceptrons (MLPs), analogous to signal-region classifiers, on the four-momenta of the final state particles along with a small set of higher-level observables to distinguish the signal from the dominant SM backgrounds. We find that the HL-LHC will be able to probe winos up to $\sim 900~$GeV, $\sim 780~$GeV, and $\sim 880~$GeV in the $Wh$ mediated $1\,\ell + 2\,b + \met$, $Wh$ mediated $1\,\ell + (\geq 2\,j) + 2\, \gamma + \met$, and $WZ$ mediated $3\ell + (\geq 2 j) + \met$ channels, respectively, for $m_{\lspone} \sim 50~$GeV, in the presence of $\lambda^{\prime \prime}_{112}u^c d^c d^c$ couplings, at $2\sigma$ sensitivity. In case the $\lambda^{\prime \prime}_{113}u^c d^c b^c$ operator is solely switched on, the projected sensitivity for winos reach up to $\sim 700~$GeV for $Wh$ mediated  $1\,\ell + (\geq 1\,b)\, + (\geq 1j)\, + 2\, \gamma + \met$ and $\sim 850~$GeV for the $WZ$ mediated $3\ell + (\geq 1 b) + \met$ channel.}

\begin{document}

\maketitle
\flushbottom

\tableofcontents 

\section{Introduction}
\label{sec:intro}

\noindent

The Standard Model~(SM) of particle physics~\cite{GLASHOW1961579,PhysRevLett.19.1264,Salam:1968rm,GELLMANN1964214} has been a highly successful theory, but it falls short in explaining several key questions, including the hierarchy problem \cite{SUSSKIND1984181,PhysRevD.14.1667}, charge-parity~(CP) asymmetry \cite{LHCb:2023tup,BaBar:2015gfp,Belle:2022uod,CDF:2014pzb,Miao:2023cvk}, the existence of dark matter~(DM) \cite{Zwicky:1933gu,1937ApJ....86..217Z,Sofue:2000jx,Jungman:1995df}, and neutrino oscillations \cite{KamLAND:2013rgu,Borexino:2013zhu,DayaBay:2018yms,Super-Kamiokande:2019gzr,NOvA:2019cyt,RENO:2018dro,T2K:2018rhz}. Supersymmetry~(SUSY) \cite{Martin:1997ns,drees2004theory,baer2006weak} provides a promising beyond the Standard Model~(BSM) theoretical framework that can address these long-standing issues, but direct experimental evidence of SUSY is yet to present itself. While the R-parity conserving~(RPC) supersymmetric scenario, which offers a stable lightest SUSY particle~(LSP) that can also serve as a viable dark matter~(DM) candidate, has been widely explored at the collider experiments, current searches impose stringent constraints on the colored sector of the RPC Minimal supersymmetric Standard Model~(MSSM), with existing bounds on squarks and gluinos above $\gtrsim 2.5~$TeV~\cite{cms_susy,atlas_susy,ATLAS:2020syg,CMS:2021beq,ATLAS:2021twp}. Its electroweakino sector remains comparatively less constrained~\cite{CMS:2019san,CMS:2020bfa,CMS:2021cox,ATLAS:2021moa,ATLAS:2021yqv,CMS:2021edw,ATLAS:2022hbt,ATLAS:2023lfr}.

An equally plausible SUSY scenario is the one with R-parity violation~(RPV), where the LSP is no longer stable and can decay into SM particles, leading to altered collider signatures. The most general RPV superpotential can be written as~\cite{Dreiner:1997uz,Banks:1995by,Barbier:2004ez,Choudhury:2024ggy}
\begin{equation} \label{eq:rpv_potential}
W_{\cancel{R}_p} = \mu_iH_u.L_i + \frac{1}{2}\lambda_{ijk}L_i.L_je_k^c + \lambda^\prime_{ijk}L_i.Q_jd_k^c + \frac{1}{2}\lambda^{\prime\prime}_{ijk}u_i^cd_j^cd_k^c,
\end{equation} 
where $H_u$, $L$, and $Q$ denote the up-type Higgs, lepton, and quark superfield doublet, respectively, and $u$, $d$, and $e$ are the up-type quark, down-type quark, and charged lepton singlet superfields, respectively. The first three operators in Eqn.~\eqref{eq:rpv_potential} parameterized by couplings $\lambda_{ijk}$ and $\lambda_{ijk}^{\prime}$ violate lepton number by one unit while the $UDD$-type operator in the last term with coupling $\lambda_{ijk}^{\prime\prime}$ is baryon number violating. Here, $i$, $j$ and $k$ are generation indices, while $\mu$ is higgsino mass term, and $c$ denotes charge conjugation. 

Our focus in the present study is on baryon number violating ($UDD$) type operators, where the LSP promptly decays into three quarks, yielding characteristically different kinematics from the RPC scenario, which typically features a large missing energy. RPV frameworks can also accommodate neutrino oscillation data~\cite{Borzumati:1996hd,Roy:1996bua,Mukhopadhyaya:1998xj,Rakshit:1998kd,Davidson:2000uc,Grossman:2003gq,Allanach:2007qc,Diaz:2014jta,Choudhury:2023lbp,Choudhury:2024yxd,Choudhury:2024ggy}, dark matter relic abundance~\cite{Takayama:2000uz,Colucci:2018yaq}, and flavour anomalies~\cite{Das:2017kfo,Domingo:2018qfg,Trifinopoulos:2019lyo,Bardhan:2021adp}. With this in mind, and given the higher statistics afforded by the upcoming High Luminosity run of the LHC~(HL-LHC), which is scheduled to operate at $\sqrt{s}=14~$TeV with integrated luminosity $\mathcal{L}=3~\textrm{ab}^{-1}$, we study the HL-LHC sensitivity to the electroweakino sector in the backdrop of two representative RPV ($UDD$) couplings: $\lambda_{112}^{\prime \prime}$ which implies $\lspone$ decaying into three light flavored jets $\lspone \to u d s$, and $\lambda_{113}^{\prime \prime}$ which implies $\lspone$ decaying into a heavy and two light flavored jets $\lspone \to u d b$. These two representative choices, $\lambda_{112}^{\prime \prime}$ and $\lambda_{113}^{\prime \prime}$, capture the collider implications for a broader class of $\lambda_{ijk}^{\prime \prime}$ couplings without the top quark index ($i=3$), which requires a separate analysis methodology \cite{Baruah:2024wrn}. We elaborate on this further in Section~\ref{sec:analysis_strategy}. 

RPV scenarios have been widely explored by ATLAS and CMS collaborations within various simplified model frameworks~\cite{CMS:2013pkf,ATLAS:2015rul,ATLAS:2015gky,CMS:2016zgb,CMS:2016vuw,CMS:2017szl,ATLAS:2018umm,CMS:2018skt,ATLAS:2019fag,ATLAS:2020wgq,CMS:2021knz,ATLAS:2021fbt}. Nonetheless, compared to the well studied electroweakino sector in R-parity conserving frameworks~\cite{Bhattacharyya:2011se,Choudhury:2012tc,Choudhury:2013jpa,Chakraborti:2014gea,Chakraborti:2015mra,Chowdhury:2016qnz,Chakraborti:2017dpu,KumarBarman:2020ylm,Barman:2022jdg,Chatterjee:2025gej}, its counterpart in the R-parity violating scenarios~\cite{Dreiner:2023bvs,Dreiner:2025kfd,Barman:2025bpx,Barman:2020azo,Bhattacherjee:2023kxw,Bhattacherjee:2013gr} remains less charted with room for further exploration. We would like to note that while the existing limits on electroweakinos in the lepton number violating $\lambda$-type RPV scenarios are typically stronger~\cite{Choudhury:2023eje,Choudhury:2023yfg}, for the baryon number violating $\lambda^{\prime \prime}$-type RPV couplings, the bounds are typically in the sub-TeV regime~\cite{CMS:2025wfw,ATLAS:2021fbt}. This prompts a detailed collider analysis to study the projected sensitivity at the upcoming HL-LHC while leveraging modern multivariate analysis techniques for efficient signal vs background discrimination to boost their discovery prospects.

We structure this paper as follows. In Section~\ref{sec:analysis_strategy}, we discuss our model framework, introduce the different types of final state configurations considered in our analysis, and outline the analysis methodology. The collider analysis strategy and results are discussed in detail in Section~\ref{sec:collider}. Finally, we conclude in Section~\ref{sec:conclusion}.

\section{Signal Processes}
\label{sec:analysis_strategy}

We consider the pair production of wino-like chargino-neutralino pairs $pp \to \lsptwo \chonepm$ at HL-LHC. Adopting a simplified framework, $\lsptwo$ and $\chonepm$ are taken to be mass-degenerate, with $\lsptwo$ subsequently decaying into $\lsptwo \to Z/h + \lspone$, and $\chonepm$ into $\chonepm \to W^{\pm} + \lspone$, where $\lspone$ is the bino-like lightest neutralino. In addition to these wino- and bino-like states, other superpartners are decoupled to a higher mass such that they do not affect our signal processes. We consider leptonic decays of the $W$ and $Z$ bosons to suppress QCD backgrounds. For the Higgs boson $h$, two scenarios are considered, one with $h \to b\bar{b}$ due to its larger branching fraction, and secondly, $h \to \gamma\gamma$, due to its cleaner signature. We introduce baryon number violating interactions through the RPV UDD $\lambda^{\prime\prime}_{ijk}$ couplings~\cite{Barbier:2004ez}, which renders $\lspone$ unstable. In Table~\ref{tab:udd_chart}, we summarize the decay modes of the $\lspone$ induced by the $\lambda^{\prime\prime}_{ijk}$ couplings. We note that all the $\lambda^{\prime\prime}$ couplings without the top quark index $i \neq 3$ can be broadly categorized into two phenomenological classes, one where the $\lspone$ decays into three light-flavored quarks, $viz.$ $\lambda^{\prime \prime}_{112}$ and $\lambda^{\prime \prime}_{212}$, and secondly, where the $\lspone$ decay involves one $b$ quark along with the light-flavored quarks, for example, $\lambda^{\prime\prime}_{113}$, $\lambda^{\prime\prime}_{123}$, $\lambda^{\prime\prime}_{213}$, and $\lambda^{\prime\prime}_{223}$. The collider signatures resulting from these two classes can be represented by the R-parity violating $\lambda^{\prime \prime}_{112}$ and $\lambda^{\prime \prime}_{113}$ couplings, and the results can be extended to other non-top quark couplings in the respective classes, without any loss of generality. On the other hand, the $\lambda^{\prime \prime}_{312}$, $\lambda^{\prime \prime}_{313}$ and $\lambda^{\prime \prime}_{323}$ couplings produce top quarks from $\lspone$ decays, resulting in altered phenomenological signatures. These top-philic couplings warrant a dedicated study of their own, which we leave for investigation in a future work~\cite{sarkarsfuture}.

\begin{table}[!htb]
\begin{center}
\begin{tabular}{|c|c|c|}
\hline
$\lambda^{\prime\prime}$ Coupling & LSP Decay & Final State \\ \hline
$\lambda^{\prime\prime}_{112}$ & $uds$ & $(W^{\pm}uds)(Zuds)$ \\ \hline
$\lambda^{\prime\prime}_{113}$ & $udb$ & $(W^{\pm}udb)(Zudb)$ \\ \hline
$\lambda^{\prime\prime}_{123}$ & $usb$ & $(W^{\pm}usb)(Zusb)$ \\ \hline
$\lambda^{\prime\prime}_{212}$ & $cds$ & $(W^{\pm}cds)(Zcds)$ \\ \hline
$\lambda^{\prime\prime}_{213}$ & $cdb$ & $(W^{\pm}cdb)(Zcdb)$ \\ \hline
$\lambda^{\prime\prime}_{223}$ & $csb$ & $(W^{\pm}csb)(Zcsb)$ \\ \hline
$\lambda^{\prime\prime}_{312}$ & $tds$ & $(W^{\pm}tds)(Ztds)$ \\ \hline
$\lambda^{\prime\prime}_{313}$ & $tdb$ & $(W^{\pm}tdb)(Ztdb)$ \\ \hline
$\lambda^{\prime\prime}_{323}$ & $tsb$ & $(W^{\pm}tsb)(Ztsb)$ \\ \hline
\end{tabular}
\caption{The complete decay modes for $pp\to\chonepm\lsptwo \to (W^{\pm}\lspone)(Z\lspone)$ channel at the HL-LHC in the presence of non-zero baryon number violating RPV coupling ($\lambda_{ijk}^{\prime\prime}$).}
\label{tab:udd_chart}
\end{center}
\end{table}

\begin{figure}[!tbh]
\centering
\begin{subfigure}{0.325\textwidth}
    \includegraphics[width=\textwidth]{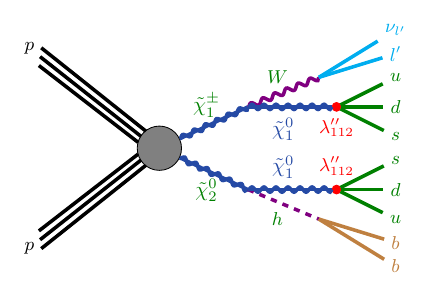}
    \caption{}
    \label{fig:first}
\end{subfigure}
\hfill
\begin{subfigure}{0.325\textwidth}
    \includegraphics[width=\textwidth]{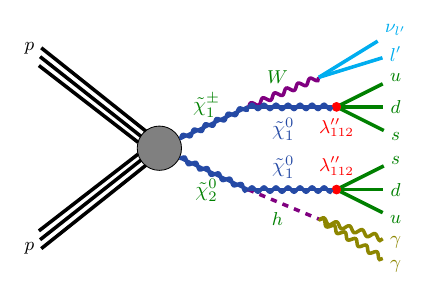}
    \caption{}
    \label{fig:second}
\end{subfigure}
\hfill
\begin{subfigure}{0.325\textwidth}
    \includegraphics[width=\textwidth]{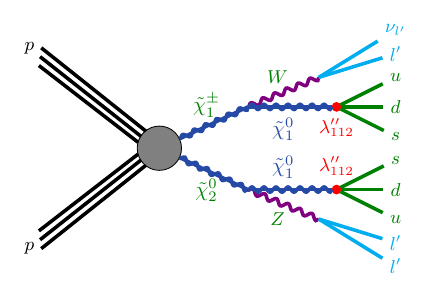}
    \caption{}
    \label{fig:third}
\end{subfigure}
\centering
\begin{subfigure}{0.325\textwidth}
    \includegraphics[width=\textwidth]{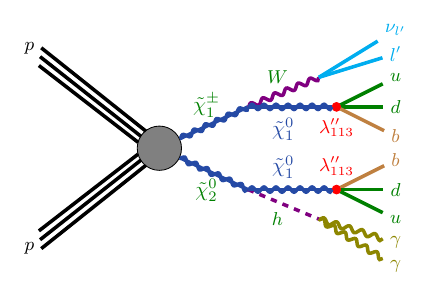}
    \caption{}
    \label{fig:sixth}
\end{subfigure}
\begin{subfigure}{0.325\textwidth}
    \includegraphics[width=\textwidth]{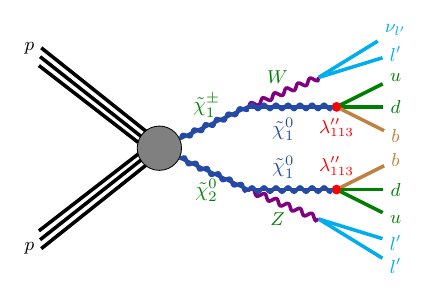}
    \caption{}
    \label{fig:fourth}
\end{subfigure}
\caption{Representative Feynman diagrams for Process (a) Process $\mathcal{P}_1$, (b) Process $\mathcal{P}_2$, (c) Process $\mathcal{P}_3$, (d) Process $\mathcal{P}_4$, and (e) Process $\mathcal{P}_5$}
\label{fig:feynman_diags}
\end{figure}

In this study, we focus our attention on the two classes of couplings, $viz.$ $\lambda^{\prime \prime}_{112}$ and $\lambda^{\prime \prime}_{113}$, which imply $\lspone \to u d s$ and $\lspone \to u d b$, respectively, and assuming that only one coupling is switched on at a time. We study three complementary final states for $\lambda^{\prime \prime}_{112}$ and two final states for $\lambda^{\prime \prime}_{113}$:

\begin{itemize}
        \item \textbf{Process $\mathcal{P}_1$~($\lambda^{\prime\prime}_{112}$)}: \\ $\lsptwo\chonepm \to (\lsptwo \to h \lspone)(\chonepm \to W^{\pm} \lspone) \to (h \to b\bar{b})\, (\lspone \to uds)\, (W^{\pm} \to \ell^{\pm} \nu)\,(\lspone \to uds)$,
        leading to the final state signature ($N_{\ell}=1$) $\cap$ ($N_b = 2$) $\cap$ $\met$. 
       
        \item \textbf{Process $\mathcal{P}_2$~($\lambda^{\prime\prime}_{112}$)}: \\ $\lsptwo\chonepm \to (\lsptwo \to h \lspone)(\chonepm \to W^{\pm} \lspone) \to (h \to \gamma\gamma)\, (\lspone \to uds)\, (W^{\pm} \to \ell^{\pm} \nu)\,(\lspone \to uds)$,
        leading to the final state signature ($N_{\ell}=1$) $\cap$ ($N_j \geq 2$) $\cap$ ($N_\gamma = 2$) $\cap$ $\met$. 

        \item \textbf{Process $\mathcal{P}_3$~($\lambda^{\prime\prime}_{112}$)}: \\ $\lsptwo\chonepm \to (\lsptwo \to Z \lspone)(\chonepm \to W^{\pm} \lspone) \to (Z \to \ell \ell)\, (\lspone \to uds)\, (W^{\pm} \to \ell^{\pm} \nu)\,(\lspone \to uds)$,
        leading to the final state signature ($N_{\ell}=3$) $\cap$ $N_j \geq 2$) $\cap$ $\met$. 

        \item \textbf{Process $\mathcal{P}_4$~($\lambda^{\prime\prime}_{113}$)}: \\ $\lsptwo\chonepm \to (\lsptwo \to h \lspone)(\chonepm \to W^{\pm} \lspone) \to (h \to \gamma \gamma)\, (\lspone \to udb)\, (W^{\pm} \to \ell^{\pm} \nu)\,(\lspone \to udb)$, leading to the final state signature ($N_{\ell}=1$) $\cap$ ($N_b \geq 1$) $\cap$ ($N_j \geq 1$) $\cap$ ($N_\gamma = 2$) $\cap$ $\met$.

        \item \textbf{Process $\mathcal{P}_5$~($\lambda^{\prime\prime}_{113}$)}: \\ $\lsptwo\chonepm \to (\lsptwo \to Z \lspone)(\chonepm \to W^{\pm} \lspone) \to (Z \to \ell \ell)\, (\lspone \to udb)\, (W^{\pm} \to \ell^{\pm} \nu)\,(\lspone \to udb)$, leading to the final state signature ($N_{\ell}=3$) $\cap$ ($N_b \geq 1$) $\cap$ $\met$


    \end{itemize}
We illustrate the tree-level Feynman diagrams for these five signal processes in Fig.~\ref{fig:feynman_diags}.
    
We utilize the \texttt{MadGraph\_aMC\@NLO}~\cite{Alwall:2014hca} framework to generate the signal and background events, with showering and hadronization performed using \texttt{Pythia8}~\cite{Sjostrand:2007gs}. Detector effects are simulated with \texttt{Delphes-3.5.0} \cite{deFavereau:2013fsa}, using the default configuration card for ATLAS. However, we set the $b$-tagging efficiency at $85\%$ and the light jet mistagging rate at $25\%$. Signal events are simulated at the leading order, while we generate two jet-matched samples for the backgrounds unless stated otherwise. 
For the signal processes, we use the next-to-leading order~(NLO)-next-to-leading-logarithmic~(NLL) cross-sections from \cite{lhc_susy}.   

We perform a machine learning based multivariate analysis utilizing fully connected Multilayer Perceptrons~(MLP) implemented within \texttt{TensorFlow}. Unlike cut-based methods, which rely on a small subset of kinematic observables, MLPs are capable of processing a much larger set of input kinematic distributions, which typically enables them to extract subtle new physics deviations with greater efficiency. We construct our neural network with 3 to 6 hidden layers, where the number of layers is optimized for each signal benchmark. The number of nodes in the hidden layers, learning rate, and batch size are optimized during training. The input and hidden layers are followed by the Rectilinear Unit~(ReLU) activation function, while Sigmoid activation is used with the output layer. Training is performed with the Adam optimizer, minimizing the sparse categorical cross-entropy loss function over 100 epochs. To avoid over-training, we implement a dropout rate of 0.1.

\section{Detailed Collider Analysis}
\label{sec:collider}
Having described the processes of our interest and the MLP setup, we take a closer look at the analysis channels in this section. We first address the  $\mathcal{P}_1,~\mathcal{P}_2$ and $\mathcal{P}_3$ channels, which are associated with the $\lambda^{\prime \prime}_{112}$ coupling, followed by the $\mathcal{P}_4$ and $\mathcal{P}_5$ channels linked to the $\lambda^{\prime \prime}_{113}$ couplings. 

\subsection{Process $\mathcal{P}_1$: ($N_{\ell}=1$) $\cap$ ($N_b = 2$) $\cap$ $\met$}
\label{sec:SR1}

We first consider the process (see Fig.~\ref{fig:first}), 
\begin{equation}
pp\to \chonepm\lsptwo \to (W\lspone)(h\lspone) \to (\ell^{\prime}\nu_{\ell^{\prime}}uds)(b\bar{b}uds).
\end{equation}
\noindent
Here, $\ell^{\prime}\equiv e,\mu,\tau$, and $\ell\equiv e,\mu$. Events are required to contain exactly one isolated lepton with $p_T > 20~$GeV and two $b$-tagged jets with $p_T > 15~$GeV. Both objects are required to lie within pseudorapidity $|\eta| < 2.5$. The invariant mass of the two $b$-tagged jets is required to lie near the mass of the Higgs boson, $m_{b_1b_2} \in [70, 130]~$GeV, where $b_1$ represents the leading $p_T$ $b$-tagged jet. After imposing these basic selection cuts, background contributions arise from $t\bar{t} + \text{jets}$, $VV + \text{jets}$~($V = Z, W$), and $Zh + \text{jets}$ processes. We choose seven representative benchmark points corresponding to different mass splittings between $\lsptwo/\chonepm$ and $\lspone$ in order to capture the different kinematic regimes, as listed in Table~\ref{tab:channel_a1_bps}. 

\begin{table}[!b]
    \centering
    \begin{tabular}{|c|c|c|} \hline 
    Benchmark Point &  $\mchonepm$ [GeV] &  $\mlspone$ [GeV] \\ \hline 
       BP1A  & 350 & 165 \\
       BP1B  & 425 & 75 \\
       BP1C  & 500 & 25 \\
       BP1D  & 500 & 100 \\
       BP1E  & 525 & 250 \\
       BP1F  & 600 & 150 \\
       BP1G  & 650 & 50 \\ \hline 
    \end{tabular}
    \caption{Benchmark points for channel $\mathcal{P}_1$ corresponding to varying masses of $\chonepm/\lsptwo$ and $\lspone$.}
    \label{tab:channel_a1_bps}
\end{table}

\begin{figure}[!t]
\begin{center}
\includegraphics[scale=0.45]{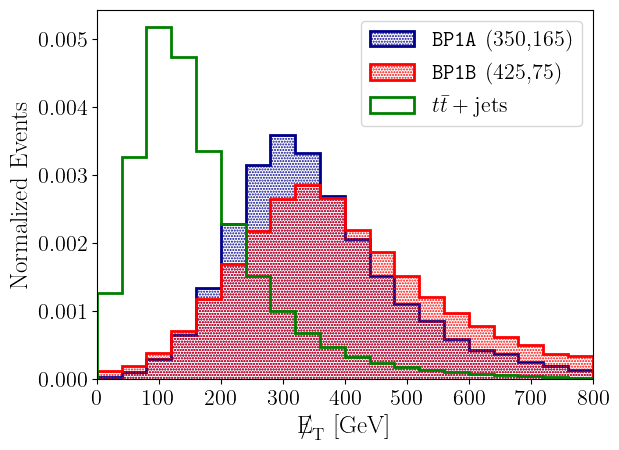} \hfill
\includegraphics[scale=0.45]{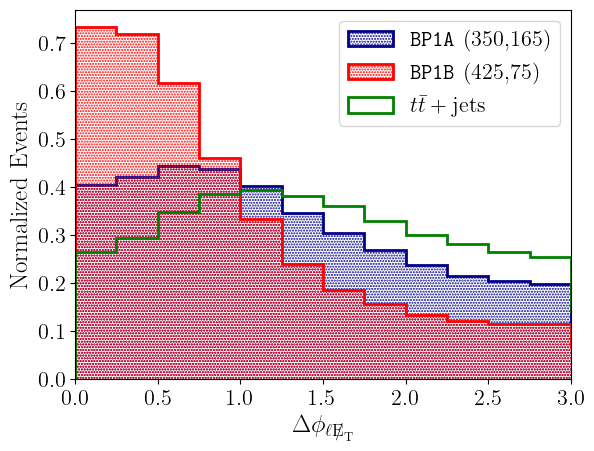}
\includegraphics[scale=0.45]{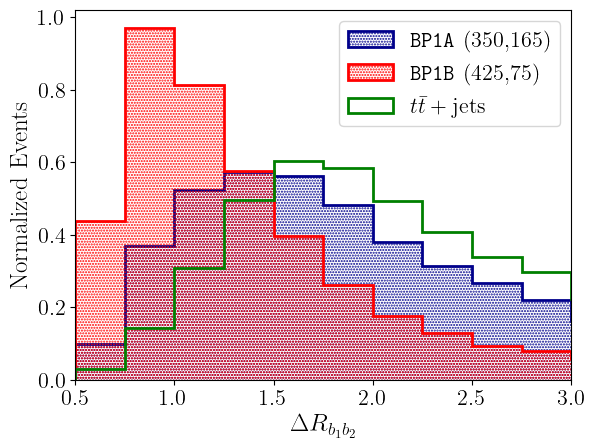} \hfill
\includegraphics[scale=0.45]{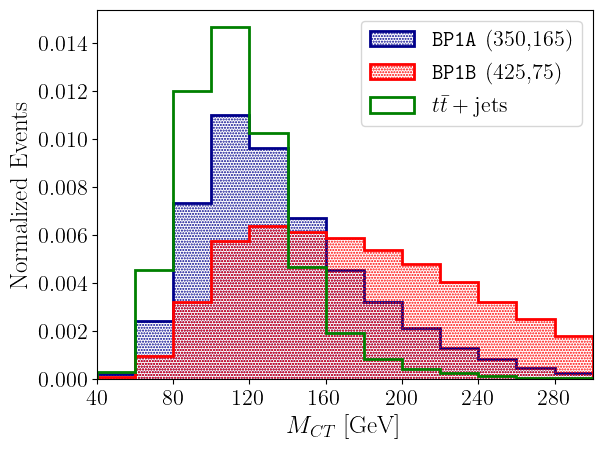}
\caption{Differential distributions for the missing transverse energy ($\met$), the contransverse mass ($M_{CT}$), $\Delta \phi$ between the lepton and missing transverse energy ($\Delta\phi_{\ell\met}$), and $\Delta R$ between the two $b$-tagged jets associated with the Higgs boson~($\Delta R_{b_1b_2}$), are shown for signal benchmark points \texttt{BP1A}$~= \{m_{\chonepm} = 350~\rm{GeV},\, m_{\lspone} = 165~\rm{GeV}\}$~(blue) and \texttt{BP1B}$~= \{m_{\chonepm} = 425~\rm{GeV},\, m_{\lspone} = 75~\rm{GeV}\}$~(red), along with the major SM background, $t\bar{t}+\rm{jets}$~(green), at the $\sqrt{s}=14~$TeV LHC.}
\label{fig:dist_SR1}
\end{center}
\end{figure}
We train the MLP classifier using the following kinematic observables:
\begin{eqnarray}
    && p_{x}^{\alpha}, p_{y}^{\alpha}, p_{z}^{\alpha}, E^\alpha \{\alpha = \ell, b_1, b_2\} \nonumber \\ 
    && \met,\, M_{CT},\,\Delta R_{b_1 b_2}, \, \Delta \phi_{\ell \met},\, N_j, \, H_T, \,M_{t_{\ell}}.
    \label{eqn:channel_a1_obs}
\end{eqnarray}
where $\{p_{x,\alpha}, p_{y,\alpha}, p_{z,\alpha}, E_\alpha\}$ represents the four-momentum of objects $\alpha$, with $\alpha$ being the final state particles like $\ell$ and $b$ jets; $\met$ represents the missing transverse energy; $M_{CT}$ is the contransverse mass, defined as \cite{Polesello:2009rn}
\begin{equation}
    M_{CT} = \sqrt{(E_T^{b_1} + E_T^{b_2})^2 - |\vec{p_T}^{b_1} - \vec{p_T}^{b_2}|^2},
\end{equation}
where $E_T^{b_{1/2}}$ and $p_{T}^{b_{1/2}}$ are the transverse momentum and energy of the respective $b$ jets; $\Delta R_{b_1 b_2}$ represents the $\Delta R = \sqrt{(\Delta \eta)^2 + (\Delta \phi)^2}$ separation between the two $b$ jets; $\Delta \phi_{\ell \met}$ is the azimuthal angle separation between $\ell$ and $\met$; $N_j$ is the number of light jets with $p_T > 15~$GeV and within $|\eta| < 2.5$; $H_T$ is the scalar sum of transverse momentum of the visible final state particles. Additionally, to suppress the massive semileptonic $t\bar{t} + \text{jets}$ background, we reconstruct the invariant mass of the leptonically decaying top quark~($t_{\ell}\to bW\to b\ell\nu$). 
Two solutions are obtained for the $z$ component of neutrino momentum ($\cancel{p}_z^{\nu}$) with $\cancel{p}_z^{\nu}=(c_1p_z^{\ell} \pm \sqrt{c_3})/c_2$, where $c_1=p_x^{\ell}\cancel{E}_x + p_y^{\ell}\cancel{E}_y$, $c_2=(E^{\ell})^2-(p_z^{\ell})^2$ and $c_3=(E^{\ell})^2c_1^2-c_2(E^{\ell})^2(\cancel{E}_x^2+\cancel{E}_y^2)$. Here $E^{\ell}$ and $p_{x,y,z}^{\ell}$ denote the energy and $x,y,z$ components of momentum of the lepton ($\ell$), respectively. 
Along with two $b$ jets, we get in total four values for $M_{\ell\met^ib^j}$ ($i,j=1,2$). Among them, the combination closest to the bare mass of the top quark is chosen. We represent the invariant mass of the aforesaid combination as ($M_{t_{\ell}}$).

We train seven different MLPs corresponding to seven signal regions~(SR1A, SR1B, SR1C, SR1D, SR1E, SR1F, and SR1G) by optimizing the signal significance metric for the 7 signal benchmarks with $\{m_{\lsptwo/\chonepm}, m_{\lspone}\}~$(in GeV): BP1A~$\{350,\ 165\}$, BP1B~$\{425,\ 75\}$, BP1C~$\{500,\ 25\}$, BP1D~$\{500,\ 100\}$, BP1E~$\{525,\ 250\}$, BP1F~$\{600,\ 150\}$ and BP1G~$\{650,\ 50\}$, respectively~(see Table~\ref{tab:channel_a1_bps}). The signal significance metric is defined as,
\begin{equation}
\sigma_{ams} = \sqrt{2\left((S+B) \times ln(1+\frac{S}{B}) - S\right)},
\end{equation}
where $S$ and $B$ are the signal and background yields at the $\sqrt{s}=14~$TeV LHC with $\mathcal{L}=3~\rm{ab}^{-1}$. 

For illustration purposes, we show the differential distributions for $\met$, $M_{CT}$, $\Delta R_{b_1 b_2}$, and $\Delta \phi_{\ell \met}$, in Fig.~\ref{fig:dist_SR1}, for two representative signal benchmarks, BP1A and BP1B, and the dominant semileptonic $t\bar{t}+jets$ background. In the SUSY signal, the primary source of missing energy is the neutrino from the leptonic $W$ boson decay. A larger mass difference between $\chonepm$ and $\lspone$ results in a boosted $W$, leading to a shift in the $\met$ spectrum towards higher values. As a result, the peak of the $\met$ distribution for BP1B~($m_{\chonepm} = 425~\rm{GeV},\, m_{\lspone} = 75~\rm{GeV}$) is further ahead at $\sim 350~$GeV, compared to that for BP1A~($m_{\chonepm} = 350~\rm{GeV},\, m_{\lspone} = 165~\rm{GeV}\}$) at around $\met \sim 300~$GeV. In the case of semileptonic $t\bar{t} + \text{jets}$ background, the missing energy primarily originates from the single neutrino in $t_{\ell} \to b (W \to \ell \nu)$, which is produced comparatively softer than the SUSY signals, due to relatively smaller cascade mass differences. This results in the $\met$ peak for semileptonic $t\bar{t}+\text{jets}$ at smaller values. A boosted $W$ boson also produces collimated $\ell$ and $\met$, yielding $\Delta \phi_{\ell \met}$ peaks at smaller values. Consequently, we observe that the $\Delta \phi_{\ell \met}$ distribution for BP1B peaks at relatively smaller values compared to BP1A. The semileptonic $t\bar{t} + \text{jets}$ background peaks further ahead with a relatively flatter distribution. Similarly, the $h$ from $\lsptwo \to h \lspone$ is more boosted for larger $\Delta m = m_{\lsptwo} - m_{\lspone}$, resulting in more collimated $b$-tagged jets produced from $h \to b\bar{b}$. This is reflected in Fig.~\ref{fig:dist_SR1}, where the $\Delta R_{b_1 b_2}$ distribution for BP1B peaks at smaller values than BP1A. In the case of semileptonic $t\bar{t} + \text{jets}$, the $b$-tagged jets are mostly uncorrelated and originate from the top and the anti-top. The $M_{CT}$ distribution is correlated with the transverse momentum of the $b$-tagged jets. The larger $\Delta m$ for BP1B results in a comparatively higher transverse momentum for the $b$ jets produced from the $h$, leading to the $M_{CT}$ distribution peaking ahead of that for BP1A. 

Signal events are simulated at different values of $m_{\lsptwo,\chonepm}$ and $m_{\lspone}$ using \texttt{Pythia-8}. $m_{\lsptwo/\chonepm}$ is varied between 150~GeV to 1~TeV with a step-size of 50~GeV, while $m_{\lspone}$ is varied from 50~GeV to $m_{\lsptwo/\chonepm} - m_h$ to ensure that the $h$ and the $W$ bosons are produced on-shell. The signal yield is computed as, 
\begin{equation}
S = \sigma_{pp \to \chonepm \lsptwo} \times Br(\chonepm \to W^{\pm}\lspone) \times Br(\lsptwo \to h \lspone) \times \mathcal{E} \times \mathcal{L},
\end{equation}
where the next-to-leading logarithm~(NNLL) and next-to-next-to-leading order~(NNLO) cross-sections for wino-like chragino-neutralino pair production are considered~\cite{lhc_susy}. $Br(\chonepm \to W^{\pm}\lspone)$,  $Br(\lsptwo \to h \lspone)$, and the successive branching fraction of $\lspone$ are assumed to be $100\%$. The signal efficiency $\mathcal{E}$ represents the fraction of signal events in the test dataset that are classified as signal-like events by the trained MLP. The background yields corresponding to the signal regions are shown in Table~\ref{tab:sigbkg_SR1}\footnote{Unless stated otherwise, we follow the same simulation pipeline, mass grids in the $\{m_{\lsptwo,\chonepm}, m_{\lspone}\}$ plane, and evaluation strategy, in all other channels analyzed in the remainder of this paper.}.
\begin{table}[!htb]
\begin{center}
\begin{tabular}{||c|c|c|c||}
\hline \hline
Benchmark Points  & Signal Yield & Background Yield & Significance \\
\hline
BP1A (350,165)  & 4429.71 & 154463.81 & 11.22 \\ \hline
BP1B (425,75)  & 1318.61 & 2648.16 & 23.85 \\ \hline
BP1C (500,25)  & 2224.98 & 14321.31 & 18.14 \\ \hline
BP1D (500,100)  & 668.58 & 1487.19 & 16.23 \\ \hline
BP1E (525,250)  & 574.08 & 4839.70 & 8.10 \\ \hline
BP1F (600,150)  & 403.40 & 2166.87 & 8.42 \\ \hline
BP1G (650,50)  & 657.35 & 5000.61 & 9.10 \\ 
\hline \hline 
\end{tabular}
\caption{The signal and background yields, and the signal significance for BP1A to BP1G as obtained from our analysis are shown. 
}
\label{tab:sigbkg_SR1}
\end{center}
\end{table}
\begin{figure}[!htb]
\begin{center}
\includegraphics[width=0.65\textwidth]{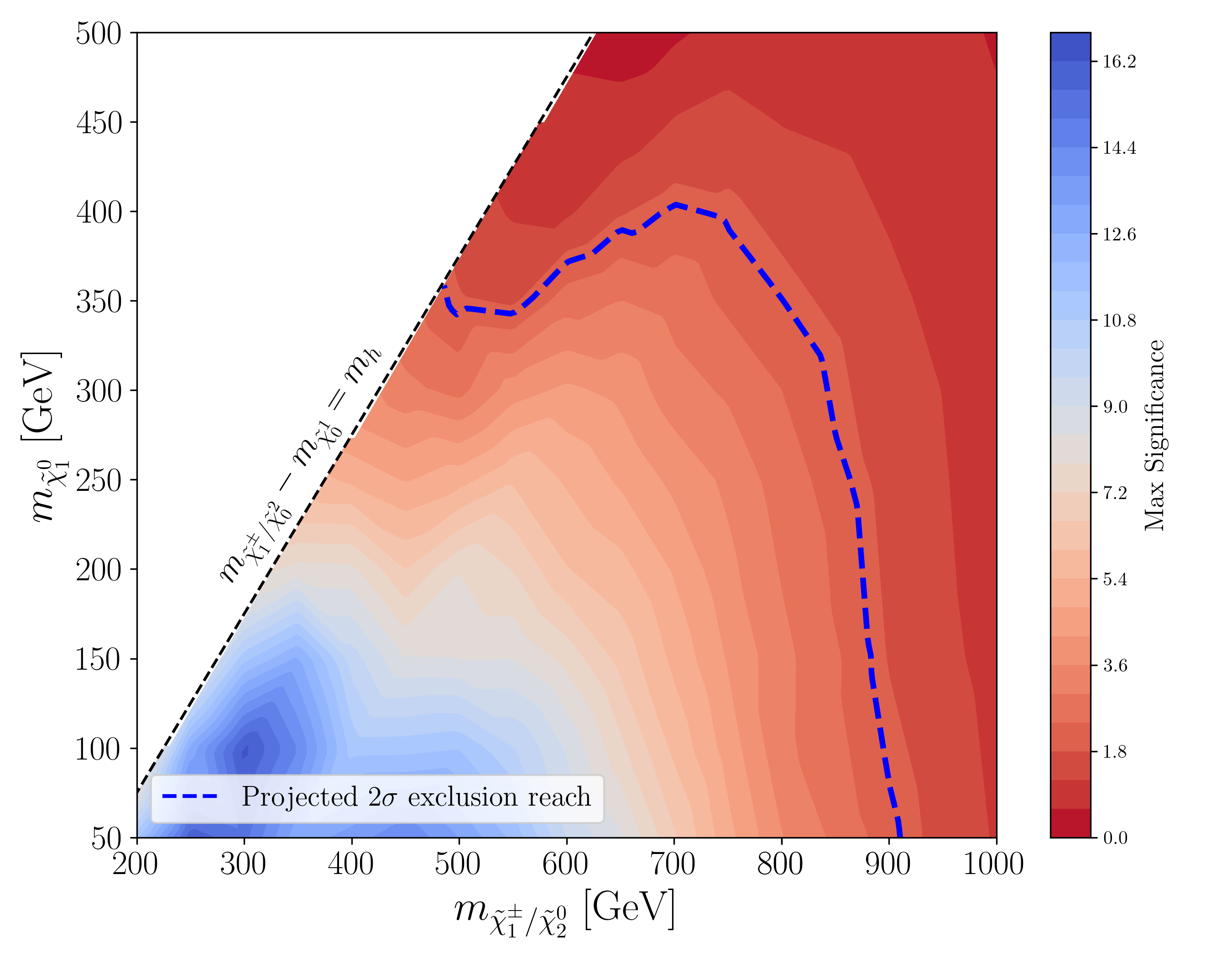}
\caption{Projected exclusion reach in the $m_{\chonepm/\lsptwo}$-$m_{\lspone}$ plane of the R-parity violating MSSM scenario with $\lambda^{\prime\prime}_{112}$ coupling via searches in the $1\ell+2b + \met$ channel at the HL-LHC is shown as a blue dashed line. The black dotted line represents the Higgs mass line where the mass difference between the NLSP and LSP is equal to the Higgs mass $m_h = 125~$GeV.}
\label{fig:exclusion_SR1}
\end{center}
\end{figure}
For every $m_{\lsptwo,\chonepm}, m_{\lspone}$, we compute the signal significances for the seven different signal regions, and the highest among them is used to derive the projection contours. In Fig.~\ref{fig:exclusion_SR1}, we show the projected $2\sigma$ sensitivity at the HL-LHC in the $m_{\lsptwo,\chonepm}, m_{\lspone}$ plane as a blue dashed line, with the signal significance values represented through the color palette. We find that within the $\lambda^{\prime\prime}_{112}$-type R-parity violating simplified scenario considered in this analysis, involving mass-degenerate wino-like $\lsptwo/\chonepm$'s and a hadronically decaying lighter bino-like $\lspone \to u d s$,  $\lsptwo,\chonepm$ masses up to $\lesssim 900~$GeV can be probed at $2\sigma$ sensitivity, for smaller $m_{\lspone} \sim 50~$GeV. The projected reach weakens as one approaches higher $m_{\lspone}$ values due to kinematic suppression, with the $2\sigma$ projection contour extending up to $m_{\lsptwo,\chonepm} \lesssim 500~$GeV at $m_{\lspone} \sim 340~$GeV.



\subsection{Process $\mathcal{P}_2$: ($N_{\ell}=1$) $\cap$ ($N_j \geq 2$) $\cap$ ($N_\gamma = 2$) $\cap$ $\met$}
\label{sec:SR2}

The signal process considered here is similar to that in process $\mathcal{P}_1$, except for the Higgs boson decaying into a pair of photons, leading to (see Fig.~\ref{fig:second})  
\begin{equation}
pp\to \chonepm\lsptwo \to (W\lspone)(h\lspone) \to (\ell^{\prime}\nu_{\ell^{\prime}}uds)(\gamma\gamma uds).
\end{equation}
Here, the events are required to contain exactly one isolated lepton and at least two light-flavored jets with $p_T > 20~$GeV, and two isolated photons with $p_T > 15~$GeV. We also impose $p_T > 20~$GeV for the leading $p_T$ photon, and $|\eta| < 2.5$ for all final state objects. Additionally, we impose that the invariant mass of the two photons must lie close to the observed mass of the Higgs boson, $120~\mathrm{GeV} < m_{\gamma\gamma} < 130~\mathrm{GeV}$, which significantly reduces the $W/Z + jets$ backgrounds. Given these basic selection cuts, major background contributions arise from $t\bar{t}h + \text{jets}$, $Wh + \text{jets}$, and $Zh + \text{jets}$. We include these three background processes in our analysis. 

\begin{table}[!htb]
    \centering
    \begin{tabular}{|c|c|c|} \hline 
    Benchmark Point &  $\mchonepm$ [GeV] &  $\mlspone$ [GeV] \\ \hline 
       BP2A  & 300 & 150 \\
       BP2B  & 425 & 100 \\
       BP2C  & 600 & 150 \\
       BP2D  & 500 & 370 \\
       BP2E  & 700 & 50 \\
       BP2F  & 650 & 300 \\
       BP2G  & 600 & 370 \\ 
       BP2H  & 550 & 250 \\ \hline 
    \end{tabular}
    \caption{Benchmark points for channel $\mathcal{P}_1$ corresponding to varying masses of $\chonepm/\lsptwo$ and $\lspone$.}
    \label{tab:channel_a2_bps}
\end{table}

To derive the projected sensitivity at the HL-LHC, we follow a similar strategy used in Sec.~\ref{sec:SR1}, selecting 8 benchmark points with small, intermediate, and large mass splittings between the wino-like $\lsptwo/\chonepm$ and bino-like $\lspone$. These benchmarks are listed in Table~\ref{tab:channel_a2_bps}. Subsequently, we train an independent MLP multi-class classifier at each benchmark, optimizing the signal significance. The classifier has four output classes, including one signal class, and three background classes, one each for $t\bar{t}h + jets$, $Wh + jets$, and $Zh + jets$. Each trained MLP classifier is considered as a `signal-region' customized to a different kinematic regime.

\begin{figure}[!htb]
\begin{center}
\includegraphics[scale=0.45]{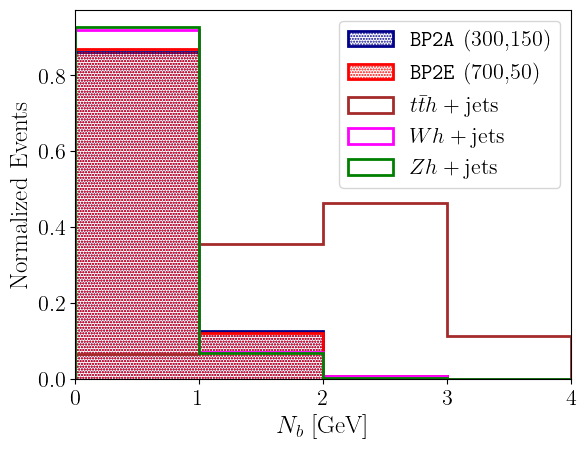} \hfill
\includegraphics[scale=0.45]{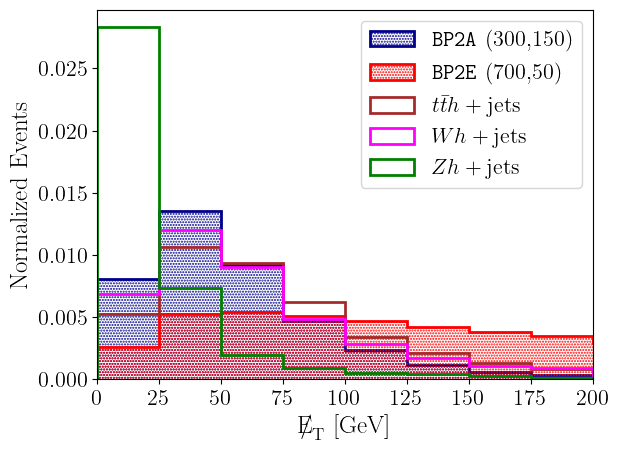}
\includegraphics[scale=0.45]{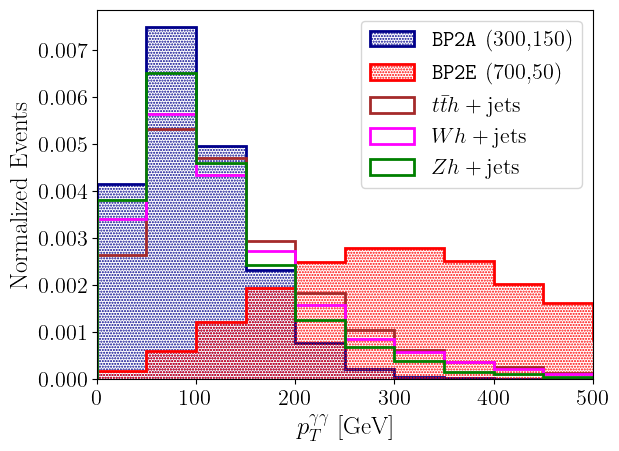} \hfill
\includegraphics[scale=0.45]{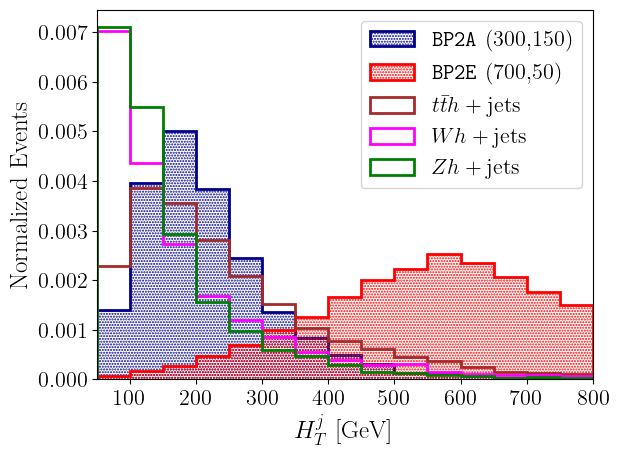}
\caption{Normalized distributions of number of bottom jets ($N_b$), missing transverse energy ($\met$), transverse momentum of di-photon system ($p_T^{\gamma\gamma}$), and scalar $p_T$ sum of light jets ($H_T^j$) are shown for two signal benchmark points, \texttt{BP2A} (300,150) and \texttt{BP2E} (700,50) and the SM backgrounds, $t\bar{t}h+\text{jets}$, $Wh+\text{jets}$, and $Zh+\text{jets}$. The signal benchmark points, \texttt{BP2A} and \texttt{BP2E}, are denoted with blue and red solid lines, respectively. The brown, magenta, and green solid lines stand for  $t\bar{t}h+\text{jets}$, $Wh+\text{jets}$, and $Zh+\text{jets}$, respectively.}
\label{fig:dist_SR2}
\end{center}
\end{figure} 

The training is performed using the following kinematic observables: 
\begin{eqnarray}
    && p_{x,\alpha}, p_{y,\alpha}, p_{z,\alpha}, E_\alpha, p_{T,\alpha} \{\alpha = \ell, \gamma_1, \gamma_2, j_1, j_2\} \nonumber \\ 
    && N_j, N_b,  \met, H_T^j, H_T^\gamma, \Delta \phi_{Wh}, M_{T}^{W\gamma_k} \{k = \gamma_1 \gamma_2\}, p_T^{\gamma_1\gamma_2}  \nonumber \\
    && \Delta R_{j_1 j_2}, \Delta R^{max}_{jj}, \Delta R^{min}_{jj}, \Delta R_{\beta \xi}, \Delta R^{min}_{\beta j}, \Delta R^{max}_{\beta j} \{\beta = h, \ell; \ \xi = j_1, j_2\},
\end{eqnarray}
where $\gamma_1$ and $\gamma_2$ represents the leading and sub-leading $p_T$ photons, respectively, and $j_1$ and $j_2$ are the leading and sub-leading light jets. $N_j$ and $N_b$ signify the number of light jets and bottom jets, respectively. $\met$ is the missing transverse energy, while $H_T^j$ and $H_T^\gamma$ represent the scalar $p_T$ sum of light jets and the two photons, respectively. $\Delta \phi_{Wh}$ denotes the azimuthal angle difference between the reconstructed $W$ and the Higgs boson $h$, where the latter is reconstructed from the diphoton system. The transverse mass of the $W$ system and the photons $\gamma_k$~($k = 1, 2$) are denoted by $M_{T}^{W\gamma_k}$, 
\begin{equation}
    M_T^{W\gamma_k} = \sqrt{(M_T^W)^2+2E_T^W E_T^{\gamma_k}-2\vec{p_T}^W\vec{p_T}^{\gamma_k}},
    \label{eq:mtwgamma}
\end{equation}
where $M_T^W$ and $E_T^W$ are the transverse mass and transverse energy of the $W$ system. We also consider the transverse momentum of the diphoton system $p_T^{\gamma_1 \gamma_2}$, and $\Delta R$ between the leading and sub-leading jets ($\Delta R_{j_1 j_2}$), the reconstructed $h$ and the light jets ($\Delta R_{h j_\xi}~\{\xi = 1,2\}$), the isolated lepton and the light jets ($\Delta R_{\ell j_\xi}$ and the smallest and highest values of both $\Delta R_{h j_\xi}$ and $\Delta R_{\ell j_\xi}$. To reconstruct the $W$ system, we attribute the missing energy $\met$ to the $W$ boson decay $W \to \ell \nu$ and impose the on-shell mass criteria, $m_W^2 = (p_\ell + p_\nu)^2 $, to reconstruct the longitudinal momentum component of the $\nu$. Among the two solutions, we choose the one where $\Delta R_{\nu\ell}$ is smaller. Events with no solutions are ignored.
In Fig.~\ref{fig:dist_SR2}, we present the differential distributions for $\met$, $p_T^{\gamma_1 \gamma_2}$ and $H_T^j$, for the signal benchmarks BP2A and BP2E, together with the relevant backgrounds $t\bar{t}h + \text{jets}$, $Wh + \text{jets}$ and $Zh + \text{jets}$. Unlike the signal process, where the missing energy predominantly arises from the $\nu$ produced in $W \to \ell \nu$ decay, in the $Zh + \text{jets}$ background, there is no explicit invisible candidate at the hard-scattering level in the relevant $Z$ decay channels~($Z \to \ell^{+}\ell^{-}$ or $Z \to q\bar{q}$), except from $\tau$ decays, jet hadronization byproducts, or `mismeasured' detector effects resulting in typically smaller $\met$ values. Consequently, $\met$ serves as a strong discriminator against the $Zh + \text{jets}$ backgrounds. A larger mass splitting $\Delta m \equiv m_{\lsptwo} - m_{\lspone}$ also means that the Higgs will be produced more boosted, shifting the $p_T^{\gamma_1\gamma_2}$ distributions. This effect is reflected in the $p_T^{\gamma\gamma}$ distributions for SR2A, which peaks at roughly $\lesssim 100~$GeV, and overlaps with the backgrounds, however, for SR2E, with a much larger $\Delta m$, $p_T^{\gamma\gamma}$ peaks around $300~$GeV. A similar trend is observed in the $H_T^j$ distributions, where larger $\Delta m$ results in more energetic jets produced from the decay of $\lspone$, shifting the $H_T^j$ distributions towards higher values.  
\begin{figure}[h]
\begin{center}
\includegraphics[width=0.65\textwidth]{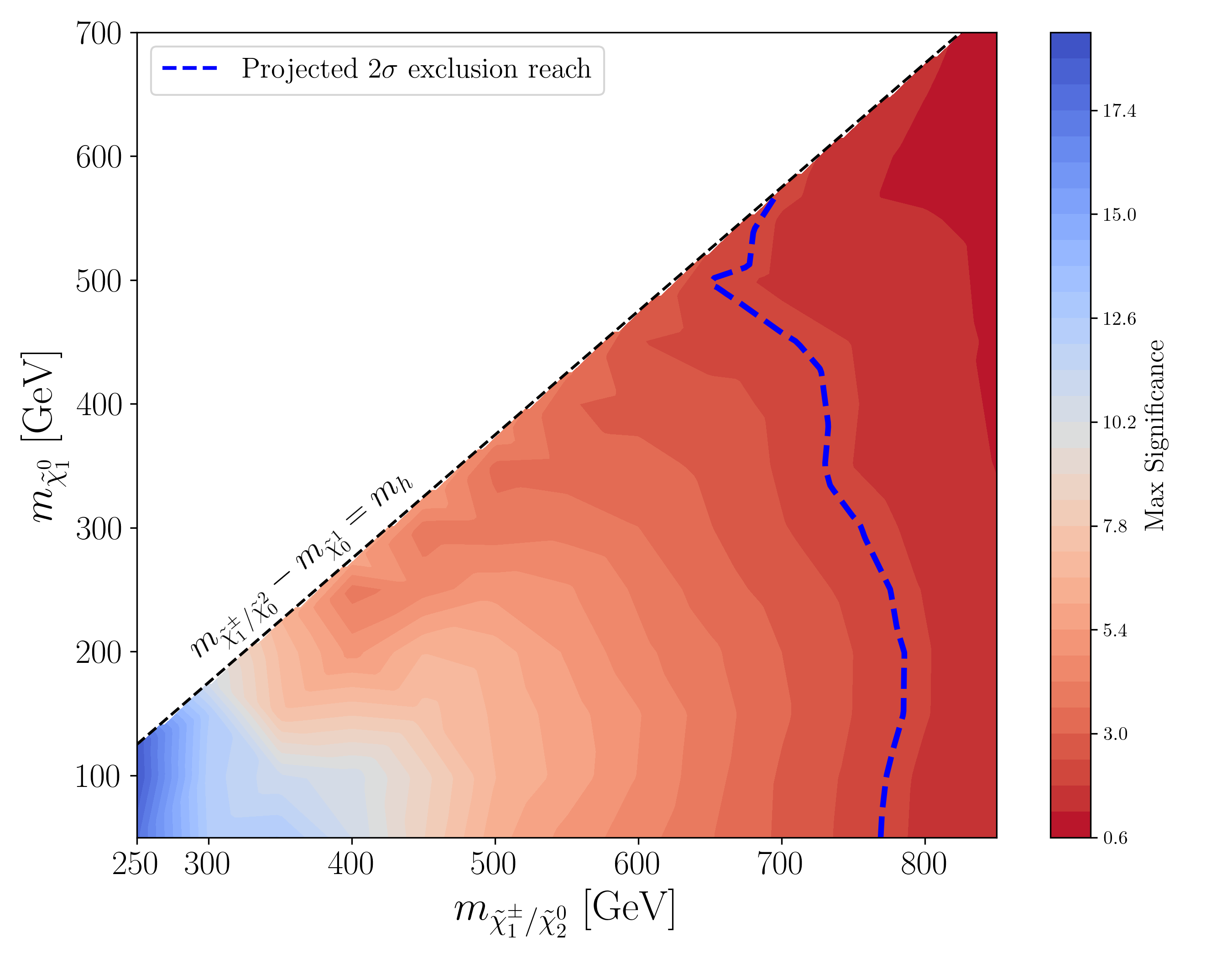}
\caption{Projected exclusion reach in the $m_{\chonepm/\lsptwo}$-$\mlspone$ plane of the $\lambda^{\prime\prime}_{112}$-type R-parity violating MSSM scenario for $1\ell+(\geq2j)+2\gamma + \met$ channel at the HL-LHC is marked with the blue dashed line. The black dashed line corresponds to $(m_{\chonepm/\lsptwo} - m_{\lspone}) = m_h$.}
\label{fig:exclusion_SR2}
\end{center}
\end{figure} 
We consider a $\{m_{\lsptwo,\chonepm}, m_{\lspone}\}$ mass grid similar to that in Section~\ref{sec:SR1}, and evaluate the signal significances at all eight `signal regions'. At each grid point, we identify the largest of the eight values, and use it to draw the $2\sigma$ projection contours at the HL-LHC in the $\{m_{\lsptwo,\chonepm}, m_{\lspone}\}$ plane, as shown in Fig.~\ref{fig:exclusion_SR2}. The color palette represents the selected highest signal significance values. 

It is observed that for smaller bino-like $\lspone$ masses~($m_{\lspone} \sim 50~$GeV), the projected $2\sigma$ reach in the wino-like $\lsptwo,\chonepm$ mass via process $\mathcal{P}_2$ extends up to $\sim 780~$GeV, which is weaker than the $2\sigma$ sensitivity via process $\mathcal{P}_1$~(analyzed in Section~\ref{sec:SR1}), which goes up to $\sim 900~$GeV. However, at higher $m_{\lspone}$, with smaller mass splittings $\Delta m$, although the Higgs is less boosted, the $\mathcal{P}_2$ channel with the $h \to \gamma \gamma$ benefits from the cleaner resonant peak, resulting in improved background discrimination. For instance, while $\{m_{\chonepm/\lsptwo} \sim 600~\textrm{GeV},\,m_{\lspone} \sim 400~$GeV$\}$ lies outside the $2\sigma$ projection contours via $\mathcal{P}_1$, it sits comfortably within the $2\sigma$ reach via $\mathcal{P}_2$. 

\subsection{Process $\mathcal{P}_3$: ($N_{\ell}=3$) $\cap$ ($N_j \geq 2$) $\cap$ $\met$}
\label{sec:SR3}
We next focus our attention on $pp \to \lsptwo/\chonepm$ production in the $WZ$ mediated cascade decay channel, with both $W$ and $Z$ decaying via leptonic modes. Just like the previous channels, we consider the R-parity violating UDD $\lambda^{\prime \prime}_{112}$ coupling, resulting in $\lspone \to uds$ decays (see Fig.~\ref{fig:third}). The full decay channel can be written as 
\begin{equation}
pp\to \chonepm\lsptwo \to (W\lspone)(Z\lspone) \to (\ell^{\prime}\nu_{\ell^{\prime}}uds)(\ell^{\prime}\ell^{\prime} uds). 
\end{equation} 
For the analysis, we select events containing exactly three isolated charged leptons ($\ell$) and at least two light jets within $|\eta| < 2.5$. The leading $p_T$ lepton ($\ell_1$) and the light jets must have $p_T > 20~$GeV, while the trailing leptons are required to have $p_T > 15~$GeV. Additionally, events must contain at least one same-flavor opposite-sign lepton pair with invariant mass $m_{\ell\ell}$ within $m_Z \pm 10~\mathrm{GeV}$, where $m_Z$ is the mass of the $Z$ boson. We also impose a $b$-veto to reduce background contributions from the $t\bar{t}~(+ X)$ processes. With these basic selection cuts, $WZ + \text{jets}$ dominates as the major background process, followed by $ZZ + \text{jets}$ and triple vector boson production~($VVV + \text{jets}$, where $V = W, Z$). 

\begin{table}[!htb]
    \centering
    \begin{tabular}{|c|c|c|} \hline 
    Benchmark Point &  $\mchonepm$ [GeV] &  $\mlspone$ [GeV] \\ \hline 
       BP3A  & 400 & 175 \\
       BP3B  & 600 & 325 \\
       BP3C  & 650 & 175 \\
       BP3D  & 550 & 400 \\
       BP3E  & 750 & 250 \\
       BP3F  & 700 & 50 \\
       BP3G  & 700 & 550 \\ 
       BP3H  & 250 & 100 \\ \hline 
    \end{tabular}
    \caption{Benchmark points for channel $\mathcal{P}_3$ corresponding to varying masses of $\chonepm/\lsptwo$ and $\lspone$.}
    \label{tab:channel_a3_bps}
\end{table}

We pursue a similar strategy as in previous sections to derive the projected reach via process $\mathcal{P}_3$ at the HL-LHC. eight signal benchmark points $\{$BP3A, BP3B, BP3C, BP3D, BP3E, BP3F, BP3G, BP3H$\}$, corresponding to different mass splittings between the wino-like $\lsptwo/\chonepm$ and bino-like $\lspone$ are considered~(see Table~\ref{tab:channel_a3_bps}).  For each benchmark, we train the MLP network on the signal events and the dominant $WZ + \text{jets}$ background, as a binary classifier, using thirty different kinematic observables,
\begin{eqnarray}
    && p_{x,\alpha}, p_{y,\alpha}, p_{z,\alpha}, E_\alpha, p_{T,\alpha} \{\alpha = \ell_1, \ell_2, \ell_3, j_1, j_2\} \nonumber \\ 
    && \met, N_j, H_T^{\ell}, H_T^j, N_{SFOS}.  \nonumber \\
    \label{eqn:sr3_training_obs}
\end{eqnarray}
In Eqn.~\eqref{eqn:sr3_training_obs}, $H_T^\ell$ and $H_T^j$ are the scalar $p_T$ sum of the charged leptons and light jets, respectively. The other observables are denoted by their usual notations. 

\begin{figure}[!htb]
    \centering
    \includegraphics[width=0.45\linewidth]{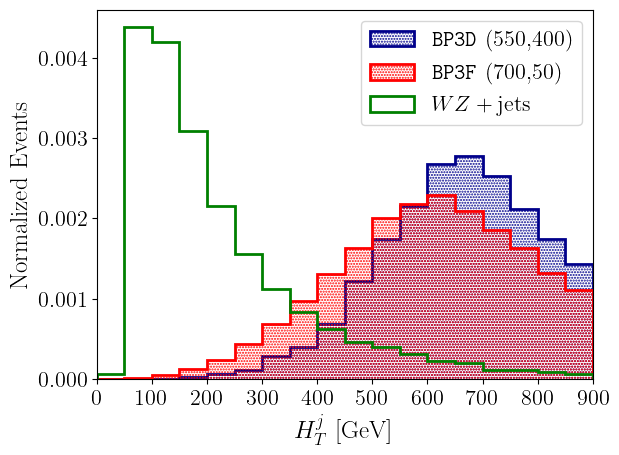}
    \includegraphics[width=0.45\linewidth]{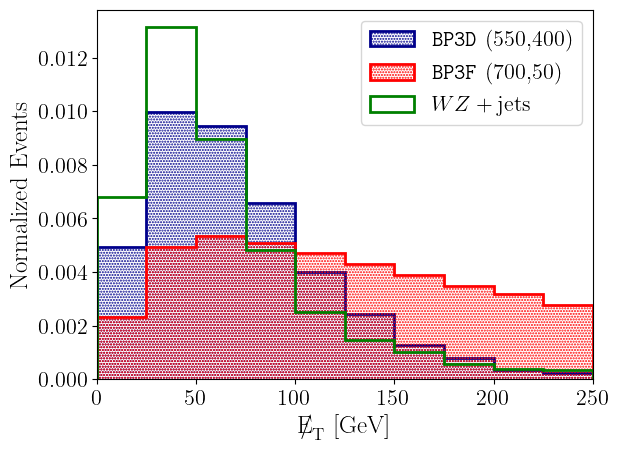}
    \includegraphics[width=0.45\linewidth]{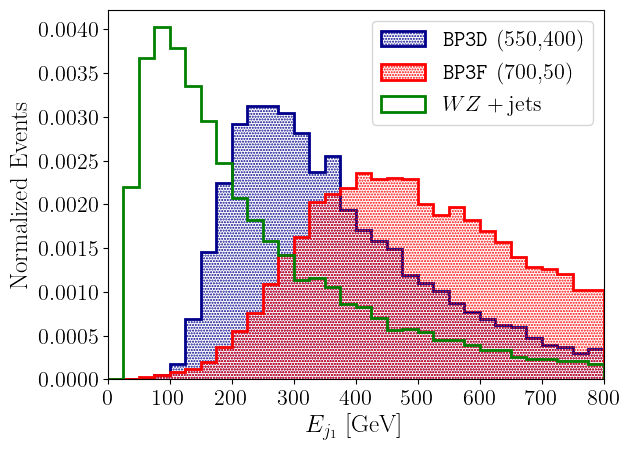}
    \caption{Normalized distributions of the scalar $p_T$ sum of light jets $H_T^j$, missing transverse energy $\met$, and the energy of the leading $p_T$ light jet $E_{j_1}$, for two representative signal benchmarks, one with intermediate mass difference $\Delta m$ between the wino-like $\chonepm/\lsptwo$ and bino-like $\lspone$ - \texttt{BP3D}$~(\{m_{\lsptwo/\chonepm},~m_{\lspone}\} = \{550, 400\}~\mathrm{GeV})$, and the other with a larger $\Delta m$, \texttt{BP3F}$~(\{m_{\lsptwo/\chonepm},~m_{\lspone}\} = \{750, 50\}~\mathrm{GeV})$, at the $\sqrt{s}=14~$TeV LHC. The distributions for the dominant $WZ + \text{jets}$ background process are also shown.}
    \label{fig:dist_SR3}
\end{figure}

In Fig.~\ref{fig:dist_SR3}, we show the distributions for $\met$, $H_T^j$, and $E_{j_1}$ for two signal benchmark points, BP3D $(\{m_{\lsptwo/\chonepm},~m_{\lspone}\} = \{550, 400\}~\mathrm{GeV})$ with intermediate $\Delta m$, and BP3F $(\{m_{\lsptwo/\chonepm},~m_{\lspone}\} = \{750, 50\}~\mathrm{GeV})$ with a larger $\Delta m$, together with the $WZ + \text{jets}$ background. In the case of BP3D, the $\lspone$ is produced with a moderate boost, resulting in the jets from its decay produced with an appreciable transverse momentum. In BP3F, owing to a larger $\Delta m$, the $\lspone$ is typically highly boosted, leading to even harder daughter jets. On the other hand, in $WZ + \text{jets}$, the light jets primarily arise from QCD radiations, which are relatively softer. This causes the $H_T^j$ distributions for both signal benchmarks peak at higher values, $H_T^j \sim 700~$GeV, while the background distribution peaks at a smaller value, roughly around $50~$GeV. It must be noted that the signal process and the $WZ + \text{jets}$ background have the same primary source for missing energy, which is the neutrino from the $W$ boson decay $W \to \ell \nu$. A larger $\Delta m$ typically boosts the $W$ boson, often resulting in more events in the tail region, as seen in its $\met$ distribution in Fig.~\ref{fig:dist_SR3}. The $\met$ distribution for BP3D, which has a smaller $\Delta m$, falls swiftly, similar to $WZ+ \text{jets}$. The LSP boost, together with how much of the LSP decay products are actually captured by the $\Delta R = 0.4$ anti-$K_T$ jet, governs the leading $p_T$ jet energy $E_{j_1}$ distribution. For the intermediate $\Delta m$ benchmark BP3D, the $\lspone$ is relatively less boosted, and the resulting jets are well separated, with the energy being evenly shared. This leads to a softer $E_{j_1}$ peak. On the other hand, in the large $\Delta m$ benchmark BP3F, the $\lspone$ has a larger boost, and the decay products are collimated. As a result, the jet reconstruction algorithm often merges two prongs into the leading $p_T$ jet, resulting in the $E_{j_1}$ distribution to peak at a higher value, $E_{j_1} \sim 450~$GeV. In the case of $WZ + \text{jets}$, the leading jet is often from QCD radiations, which are typically softer, with its peak at a much smaller value $E_{j_1} \sim 100~$GeV. 
\begin{figure}[!htb]
    \centering
    \includegraphics[width=0.65\linewidth]{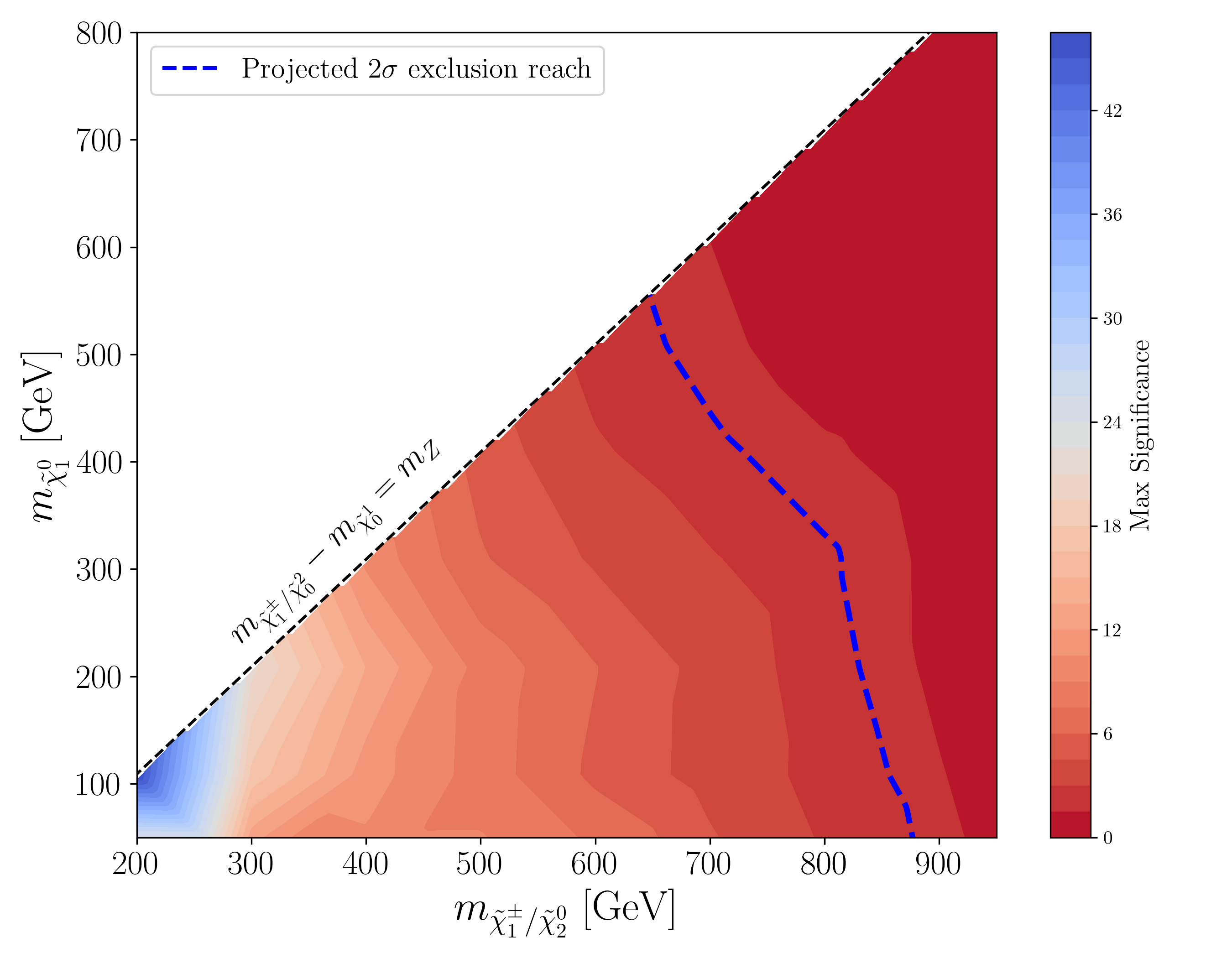}
    \caption{Projected sensitivity in the $m_{\chonepm/\lsptwo}$-$\mlspone$ plane from searches in the $3\ell+(\geq2j)$ + $\met$ channel in the $\lambda^{\prime\prime}_{112}$-type R-parity violating MSSM scenario at the $\sqrt{s}=14~$TeV LHC with $\mathcal{L}= 3~\mathrm{ab}^{-1}$. The $2\sigma$ contour is shown as a blue dashed line. The signal significance values are represented via the color palette.}
    \label{fig:exclusion_SR3}
\end{figure}
We derive the HL-LHC projection contours in the $m_{\lsptwo/\chonepm}$-$m_{\lspone}$ plane following a similar strategy adopted in Secs.~\ref{sec:SR1} and \ref{sec:SR2}. Our results are presented in Fig.~\ref{fig:exclusion_SR3}. We note that for training, while we use a binary classifier~(signal vs. the dominant background $WZ + jets$), when evaluating the projected sensitivities, we consider a test dataset which includes the dominant $WZ + jets$ and other sub-dominant backgrounds $viz.$ $ZZ + jets$ and $VVV + jets$ as well. We find that wino-like $\lsptwo/\chonepm$ masses up to $\sim 880~$GeV can be probed at the HL-LHC for small bino-like $\lspone$ masses, $m_{\lspone} \sim 50~$GeV, at $2\sigma$ sensitivity. For larger $m_{\lspone} 
\sim 500~$GeV, the projected $2\sigma$ reach extends up to around $m_{\lsptwo/\chonepm} \sim 640~$GeV.
\subsection{Process $\mathcal{P}_4$: ($N_{\ell}=1$) $\cap$ ($N_b \geq 1$) $\cap$ ($N_j \geq 1$) $\cap$ ($N_\gamma = 2$) $\cap$ $\met$}
\label{sec:SR4}
We now shift our attention to the R-parity violating $\lambda^{\prime\prime}_{113}$ coupling, in which case, $\lspone$ promptly decays via $\lspone \to u d b$, leading to a hadronically rich final state. We first consider the wino-like chargino-neutralino pair production, followed by their cascade decay via the $W$ and $h$, respectively, with the Higgs boson decaying into a diphoton system $ h \to \gamma\gamma$ and the $W$ boson decaying leptonically $W \to \ell \nu$. The cascade decay chain is similar to process $\mathcal{P}_2$, but with the exception of $\lspone$ decaying via $\lspone \to udb$ (see Fig.~\ref{fig:sixth}). The complete decay channel is as follows:
\begin{equation}
pp\to \chonepm\lsptwo \to (W\lspone)(h\lspone) \to (\ell^{\prime}\nu_{\ell^{\prime}}udb)(\gamma\gamma udb)
\end{equation} 
We select events containing exactly one isolated lepton, exactly two photons, and at least one $b$-tagged jet and one light jet, within $|\eta| < 2.5$. The leading $p_T$ lepton and photon are required to have $p_T > 20~$GeV, while other final state objects must have $p_T > 15~$GeV. We also require the events to have the di-photon invariant mass lying within the range $120~\mathrm{GeV} \leq m_{\gamma\gamma} \leq 130~\mathrm{GeV}$. In this channel, the dominant background is $t\bar{t}h + \text{jets}$, while sub-leading contributions arise from $t\bar{t}\gamma\gamma + \text{jets}$, $Wh + \text{jets}$ and $Zh + \text{jets}$. 

\begin{figure}[!htb]
    \centering
    \includegraphics[width=0.45\linewidth]{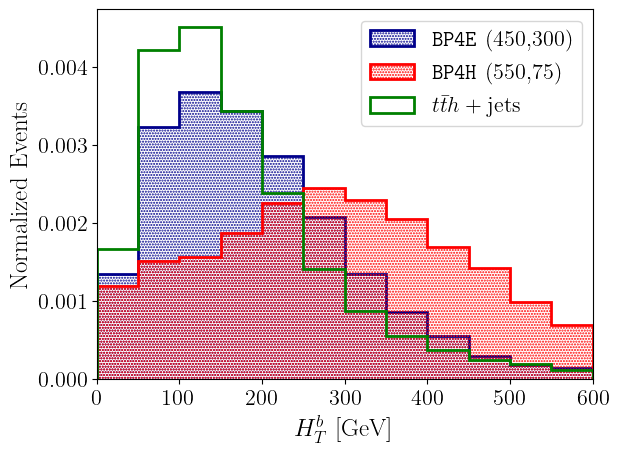}
    \includegraphics[width=0.45\linewidth]{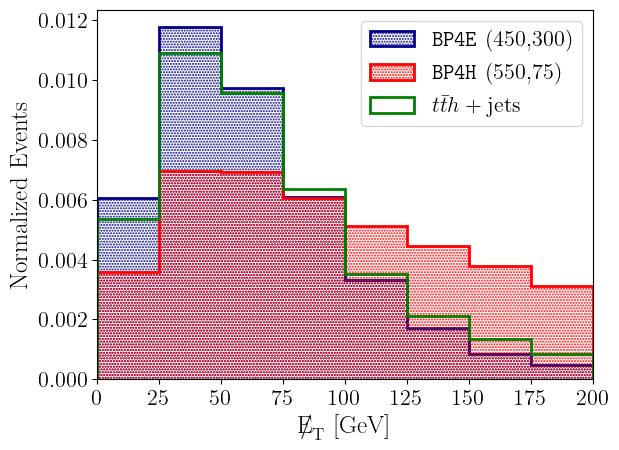}
    \includegraphics[width=0.45\linewidth]{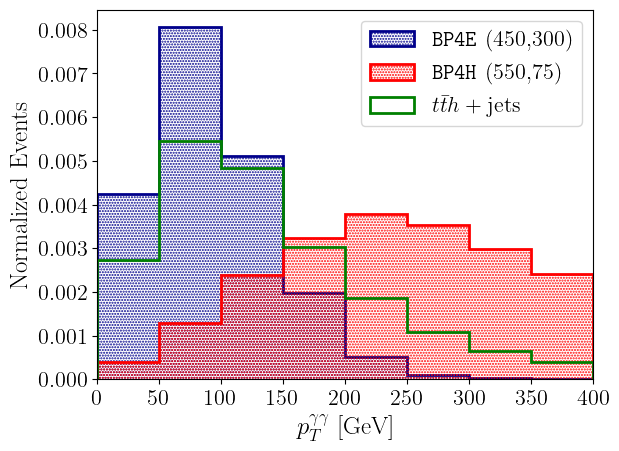}
    \caption{Normalized distributions of the scalar $p_T$ sum of the $b$-tagged jets $H_T^b$, missing transverse energy $\met$, and the transverse momentum of the diphoton system $p_T^{\gamma\gamma}$, for two representative signal benchmark points, BP4E$~(\{m_{\lsptwo/\chonepm},~m_{\lspone}\} = \{450, 300\}~\mathrm{GeV})$ and BP4H$~\{550, 75\}~\mathrm{GeV}$, and the $t\bar{t}h + \text{jets}$ background, at the $\sqrt{s}=14~$TeV LHC.}
    \label{fig:dist_SR4}
\end{figure}

\begin{table}[!htb]
    \centering
    \begin{tabular}{|c|c|c|} \hline 
    Benchmark Point &  $\mchonepm$ [GeV] &  $\mlspone$ [GeV] \\ \hline 
       BP4A  & 250 & 100 \\
       BP4B  & 400 & 200 \\
       BP4C  & 400 & 250 \\
       BP4D  & 425 & 250 \\
       BP4E  & 450 & 300 \\
       BP4F  & 500 & 250 \\
       BP4G  & 525 & 150 \\ 
       BP4H  & 550 & 75 \\ 
       BP4I  & 600 & 450 \\ \hline
    \end{tabular}
    \caption{Benchmark points for channel $\mathcal{P}_4$ corresponding to varying masses of $\chonepm/\lsptwo$ and $\lspone$.}
    \label{tab:channel_a4_bps}
\end{table}
Following a similar strategy to the previous analyses, we select nine signal benchmark points (BP4A-BP4I) with small, intermediate, and large $\Delta m \equiv (m_{\chonepm/\lsptwo} - m_{\lspone})$. The benchmark points are listed in Table~\ref{tab:channel_a4_bps}. We train nine independent MLP networks as a binary classifier for the signal and the dominant $t\bar{t}h + \text{jets}$ background. The kinematic observables used to perform the training are, 
\begin{eqnarray}
    && p_{x,\alpha}, p_{y,\alpha}, p_{z,\alpha}, E_\alpha, p_{T,\alpha} \{\alpha = \ell, \gamma_1, \gamma_2, b_1, j_1\} \nonumber \\ 
    && \met, H_T^{\gamma}, H_T^j, H_T^b, p_T^{\gamma\gamma}, \Delta\phi_{Wh}, M_T^{W\gamma_k} ~~\text{where}~~k=1,2 \nonumber \\
    && \Delta R_{\ell, b_1}, \Delta R_{h,b_1}, \Delta R_{\ell,h}, \Delta R_{\delta,\xi}^{\text{max}}, \Delta R_{\delta,\xi}^{\text{min}} ~\{\delta = \ell, \gamma; \xi = b,j\}, \nonumber
    \label{eqn:kinematic_obs_SR5}
\end{eqnarray} 
where the notations have their usual meaning. For illustration, in Fig.~\ref{fig:dist_SR4}, we show the distributions for the scalar $p_T$ sum of the $b$-tagged jets $H_T^b$, missing transverse energy $\met$, and the transverse momentum of the diphoton system $p_T^{\gamma\gamma}$, for the signal benchmark points BP4E and BP4H, and the leading background - $t\bar{t}h + \text{jets}$.

In the signal process, the $b$-tagged jets are primarily produced from $\lspone$ decay. As such, their scalar $p_T$ sum~($H_T^b$) scales with the boost of the $\lspone$, which, in turn, is governed by the mass difference $\Delta m$. For BP4H, which has a larger $\Delta m$, the resulting $H_T^b$ distribution peaks at a higher value $H_T^b \sim 300~$GeV than in the case of BP4E, which has a smaller mass splitting. In $t\bar{t}h + jets$, $H_T^b$ is dominated by $b$-tagged jets from top quark decays. It is observed that its $H_T^b$ distribution peaks at a smaller value than BP4H, but overlaps with the tail region of BP4E. The missing energy distributions, which are largely governed by the only neutrino produced from $W \to \ell \nu$ decay in both signal and the background, exhibit a comparable shape between the smaller $\Delta m$ signal process from BP4E and the $t\bar{t}h + \text{jets}$ background. However, in the case of BP4H, the larger mass splitting between $\chonepm$ and $\lspone$ leads to a boosted $W$, which results in a flatter $\met$ tail, while the peak is retained at similar values to that for the $t\bar{t}h + \text{jets}$ background. Similar to $H_T^b$, the transverse momentum of the di-photon system $p_T^{\gamma\gamma}$ is also correlated with the mass difference between $\lsptwo$ and $\lspone$. For the signal benchmark BP4H with a larger $\Delta m$, the $p_T^{\gamma\gamma}$ distribution displays a noticeable shift towards higher values, with its peak at $p_T^{\gamma\gamma} \sim 225~$GeV, while for BP4E, the peak roughly coincides with $t\bar{t}h + \text{jets}$, but with a softer tail.

\begin{figure}[!t]
    \centering
    \includegraphics[width=0.65\linewidth]{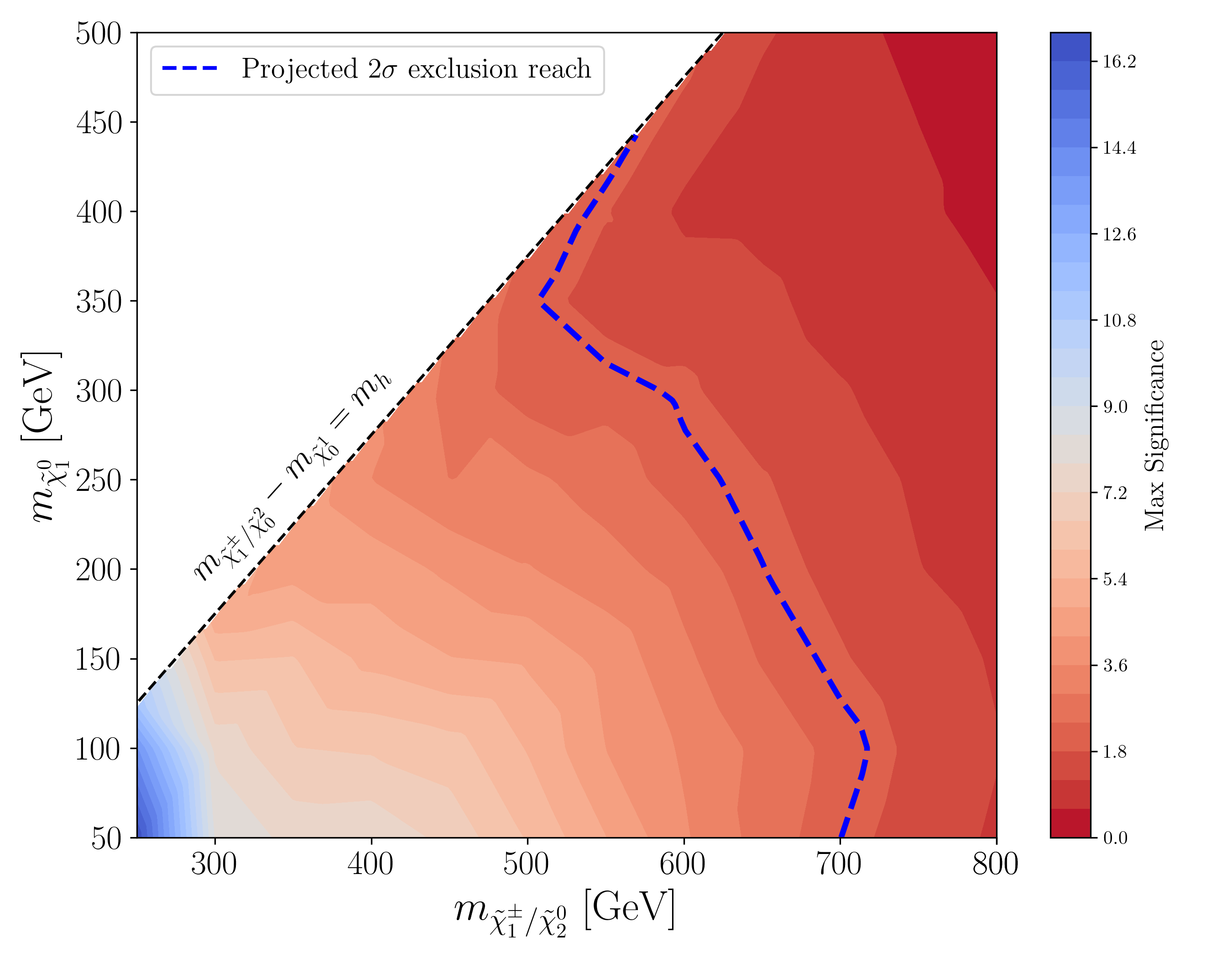}
    \caption{Projected sensitivity in the $m_{\chonepm\lsptwo}-m_{\lspone}$ plane, considering the $\lambda^{\prime\prime}_{113}$-type R-parity violating MSSM scenario, from searches in the $1\ell+(\geq 1b)+(\geq 1j)+2\gamma+\met$ channel at the HL-LHC. The $2\sigma$ reach is shown as a blue dashed line. The color palette represents the signal significance values.}
    \label{fig:exclusion_SR4}
\end{figure}

We train nine independent MLP classifiers to distinguish the signal process and the dominant $t\bar{t}h + \text{jets}$ background. Following the notation and labels from previous analyses, we show the projected $2\sigma$ sensitivity at the HL-LHC in the $m_{\chonepm/\lsptwo}$-$m_{\lspone}$ plane in Fig.~\ref{fig:exclusion_SR4} as a blue dashed line. The signal significance values at the HL-LHC are also shown as a color palette. In the $pp \to 1\ell + 2\gamma + (\geq 1 j)  + (\geq 1 b)$ channel considered above, with the R-parity violating $\lambda^{\prime \prime}_{113}$ coupling, wino-like $\chonepm/\lspone$ can be probed up to $m_{\chonepm/\lsptwo}\sim 700~$GeV for a lighter bino-like $\lspone$ with mass $m_{\lspone} \sim 50~$GeV, while the projected reach for the wino-like electroweakinos shrinks to $m_{\chonepm/\lsptwo} \sim 500~$GeV, for $m_{\lspone} \sim 350~GeV$.

\subsection{Process $\mathcal{P}_5$: ($N_{\ell}=3$) $\cap$ ($N_b \geq 1$) $\cap$ $\met$}
\label{sec:SR5}
We finally study the HL-LHC prospects for the $WZ$ mediated cascade decay of wino-like chargino-neutralino pair production $pp \to \chonepm \lsptwo$ in the $\lambda^{\prime\prime}_{113}$ R-parity violating scenario, with both the $W$ and $Z$ boson decaying leptonically~(see Fig.~\ref{fig:fourth}). The decay channel is 
\begin{equation}
pp\to \chonepm\lsptwo \to (W\lspone)(Z\lspone) \to (\ell^{\prime}\nu_{\ell^{\prime}}udb)(\ell^{\prime}\ell^{\prime} udb)
\end{equation} 

\begin{table}[!t]
    \centering
    \begin{tabular}{|c|c|c|} \hline 
    Benchmark Point &  $\mchonepm$ [GeV] &  $\mlspone$ [GeV] \\ \hline 
       BP5A & 250 & 135 \\
       BP5B & 600 & 205 \\
       BP5C & 350 & 100 \\
       BP5D & 450 & 300 \\
       BP5E & 500 & 200 \\
       BP5F & 550 & 350 \\
       BP5G & 680 & 250 \\ 
       BP5H & 600 & 50 \\ \hline 
    \end{tabular}
    \caption{Benchmark points for channel $\mathcal{P}_5$ corresponding to varying masses of $\chonepm/\lsptwo$ and $\lspone$.}
    \label{tab:channel_a5_bps}
\end{table}

For our collider analysis, we select events containing exactly three isolated charged leptons with $p_T > 15~$GeV, and at least one same-flavor opposite-sign lepton pair with invariant mass $m_{\ell\ell}$ within $m_Z \pm 10~$GeV. The leading $p_T$ lepton is also required to have $p_T > 20~$GeV. Additionally, the event must have at least one $b$-tagged jet with $p_T > 15~$GeV. The charged leptons and $b$-tagged jets are also required to fall within $|\eta| < 2.5$. We note that the major background for this channel arises from $WZ + jets$ and $t\bar{t} + \text{jets}$, while subdominant contributions arise from $t\bar{t}Z + \text{jets}$, $WZZ + \text{jets}$, and $ZZZ + \text{jets}$. 

\begin{figure}[!b]
    \centering
    \includegraphics[width=0.45\linewidth]{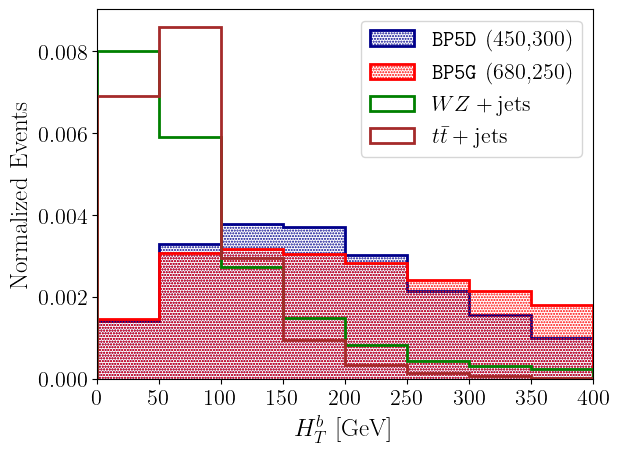}
    \includegraphics[width=0.45\linewidth]{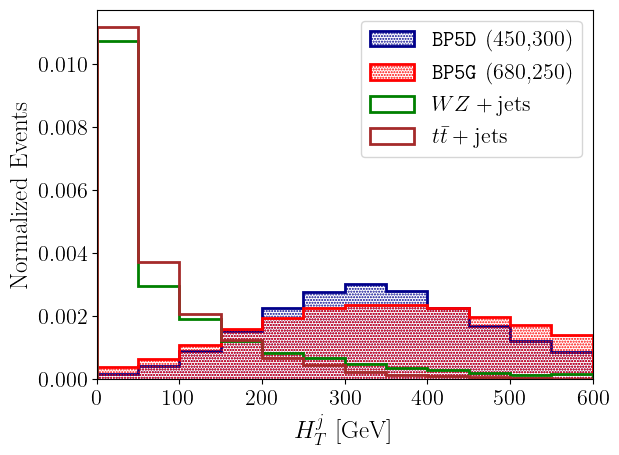}
    \includegraphics[width=0.45\linewidth]{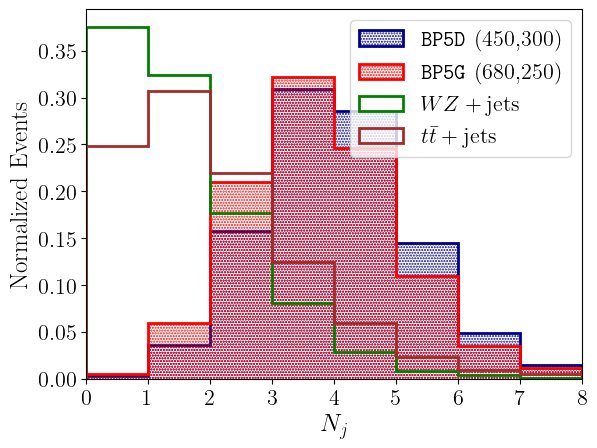}
    \caption{Normalized distributions of the scalar $p_T$ sum of the $b$-tagged jets $H_T^b$, scale $p_T$ sum of the light jets $H_T^j$, and the number of light jets $N_j$, for benchmark points BP5D and BP5G, and the major backgrounds, $WZ + \text{jets}$ and $t\bar{t} + \text{jets}$, for process $\mathcal{P}_5$, at the $\sqrt{s}=14~$TeV LHC with $\mathcal{L} = 3~\textrm{ab}^{-1}$.}
    \label{fig:dist_SR5}
\end{figure}

We select eight benchmark points with small, intermediate, and large mass splittings between the wino-like and bino-like electroweakinos, as listed in Table~\ref{tab:channel_a5_bps}. Accordingly, eight independent MLP multi-class classifiers are trained to categorize the signal and the two major backgrounds, $WZ + \text{jets}$ and $t\bar{t} + \text{jets}$. The training is performed using the following kinematic observables, 
\begin{eqnarray}
    && p_{x,\alpha}, p_{y,\alpha}, p_{z,\alpha}, E_\alpha, p_{T,\alpha} \{\alpha = \ell_1, \ell_2, \ell_3, b_1\} \nonumber \\ 
    && \met, N_b, N_j, H_T^{\ell}, H_T^j, H_T^b, N_{SFOS}  \nonumber
\end{eqnarray}
where the notations have their usual meaning. 

In Fig.~\ref{fig:dist_SR5}, we illustrate the normalized distributions for three of these observables, the scalar $p_T$ sum of the $b$ tagged jets $H_T^b$, the scalar $p_T$ sum of the light jets $H_T^j$ and the number of light jets $N_j$, for two signal benchmarks BP5D and BP5G, and the two dominant backgrounds, $WZ + \text{jets}$ and $t\bar{t} + \text{jets}$. The $H_T^b$ and $H_T^j$ distributions for both signal benchmarks BP5D and BP5G peak at similar values ahead of the backgrounds. While the $\lspone$ is typically more boosted in BP5G than in BP5D, the corresponding gain in $H_T^b$ and $H_T^j$ is mostly offset due to more collimated decay products from $\lspone$ in the former case, which can merge the light jet constituents into the reconstructed $b$ jet, leading to slightly reduced $b$-tagging rates. However, the larger $\Delta m$ does lead to harder jets, which results in more events populating the high $H_T$ regions. Accordingly, the tails of both $H_T^b$ and $H_T^j$ fall more slowly in the case of BP5G than for BP5D. It is also observed that the signal process involves a larger number of light jets in the final state, as compared to $WZ + \text{jets}$ and $t\bar{t} + \text{jets}$, offering a good discrimination between signal and backgrounds. 

\begin{figure}[!t]
    \centering
    \includegraphics[width=0.65\linewidth]{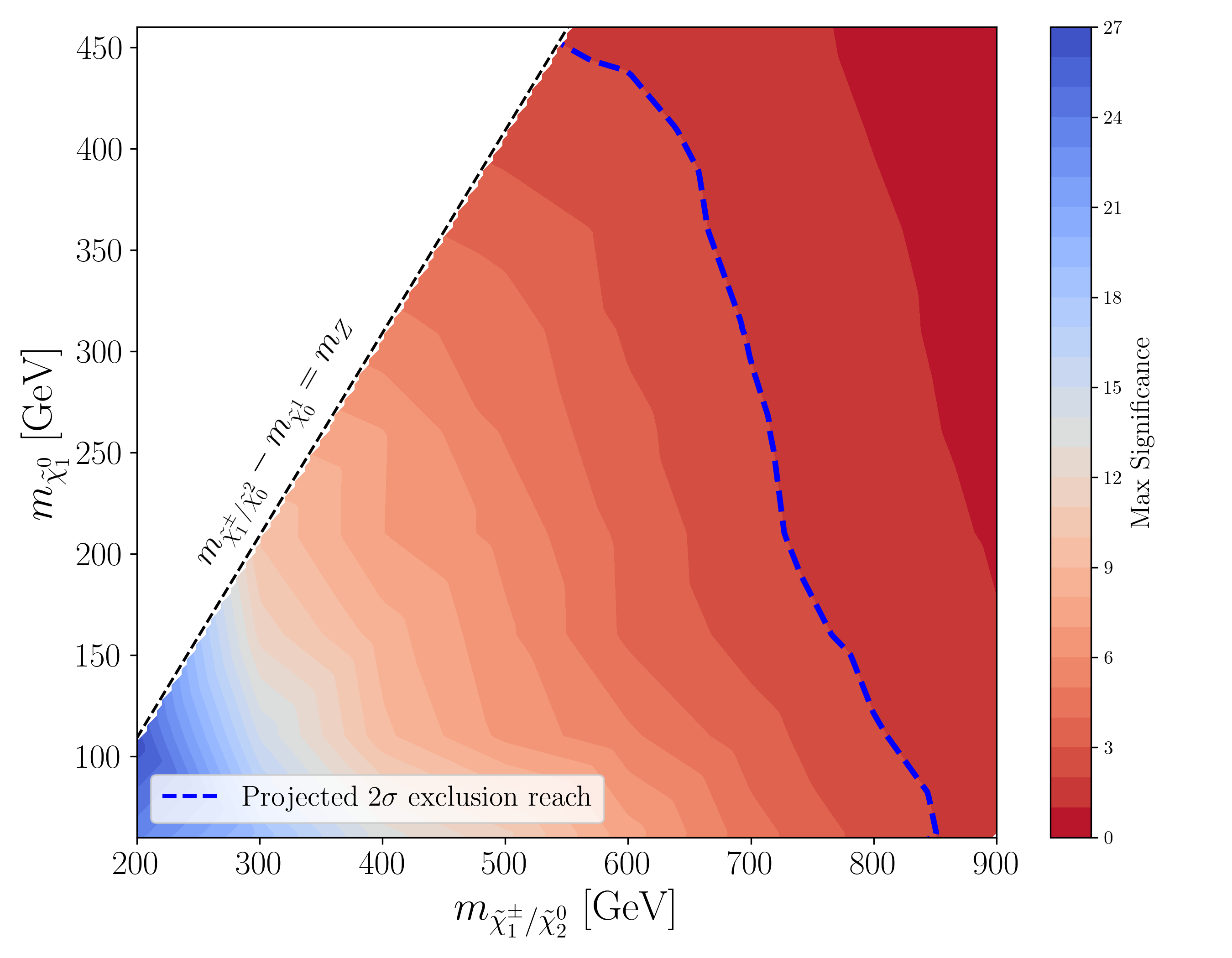}
    \caption{Projected sensitivity in the $m_{\chonepm\lsptwo}-m_{\lspone}$ plane, considering the $\lambda^{\prime\prime}_{113}$-type R-parity violating MSSM scenario, from searches in the $3\ell+(\geq1b) + \met$ channel at the HL-LHC. The $2\sigma$ reach is shown as a blue dashed line. Signal significance values are represented through the color palette.}
    \label{fig:exclusion_SR5}
\end{figure}

We present the projected reach at the HL-LHC in the $m_{\chonepm\lsptwo}-m_{\lspone}$ plane in Fig.~\ref{fig:exclusion_SR5}, with the $2\sigma$ contours shown as a blue dashed line. We show the signal significance values as a color palette. It is observed that for low bino-like $\lspone$ masses,  $\mlspone \sim 50~$GeV, wino-like $\chonepm/\lsptwo$ can be probed up to roughly $\sim 850~$GeV, at $2\sigma$ sensitivity. The projected reach declines to roughly $m_{\chonepm/\lsptwo} \sim 550~$GeV, for a heavier $\lspone$ with mass at roughly $\sim 450~$GeV.   
\section{Summary and Conclusion}
\label{sec:conclusion}

The current LHC data is yet to reveal any significant deviation from SM predictions, placing increasingly strong constraints on the various BSM scenarios. In this context, the electroweak sector of the baryon number violating supersymmetric scenario remains an attractive framework to be pursued at the high luminosity LHC, given that the existing bounds on this framework are still below the $\mathcal{O}(1)~$TeV scale. With modern machine learning techniques in place that are capable of learning from the complex high-dimensional collider data, enabling improved signal-background categorization, it is of interest to reassess the future reach of this sector at the HL-LHC, complementing the traditional analysis strategies.

In this work, we have considered a simplified model framework featuring roughly mass degenerate wino-like NLSPs, $\chonepm$ and $\lsptwo$, and a bino-like LSP, $\lspone$ that decays hadronically owing to the baryon number violating $\lambda_{112}^{\prime\prime}$ and $\lambda_{113}^{\prime\prime}$  couplings,~(with only one coupling switched on at a time). We have investigated the HL-LHC prospects of direct wino production $pp \to \chonepm \lsptwo$.  The wino-like NLSPs ($\chonepm/\lsptwo$) decay to the bino-like LSP ($\lspone$) and $W~\text{and}~Z/h$, and the LSP further decays to SM quarks via the aforementioned RPV couplings. We have studied the collider implications of $\lambda_{112}^{\prime\prime}$ coupling in the $Wh$ mediated $1\ell+2b+\met$, $Wh$ mediated $1\ell+(\geq 2j)+2\gamma + \met$, and $WZ$ mediated $3\ell+(\geq 2j)+\met$ channels, discussed in Section~\ref{sec:SR1}, Section~\ref{sec:SR2}, and Section~\ref{sec:SR3}, respectively. We further delve into the collider phenomenology of  $\lambda_{113}^{\prime\prime}$ coupling on direct wino production searches in $Wh$ mediated $1\ell + (\geq 1b)+(\geq 1j)+2\gamma+\met$ channel and $WZ$ mediated $3\ell+(\geq 1b)+\met$ channel, as discussed in Section~\ref{sec:SR4} and Section~\ref{sec:SR5}, respectively. We adopt MLP-based classifiers to perform signal-background discrimination in all these channels. For every channel, a set of benchmark-specific MLPs are trained, using a combination of four-momenta and higher-level kinematic variables.

Our analysis indicates that wino masses can be excluded up to $\sim 900$ GeV for a bino mass of $\sim$50 GeV at $2\sigma$ sensitivity at the HL-LHC in $Wh$ mediated $1\ell+2b+\met$ channel. The projected $2\sigma$ exclusion reach for the same in $Wh$ mediated $1\ell+(\geq 2j)+2\gamma + \met$ and $WZ$ mediated $3\ell+(\geq 2j)+\met$ channels are obtained to be $\sim 780$ GeV and $\sim880$ GeV, respectively. Moreover, the projected $2\sigma$ exclusion reach for $m_{\chonepm=\lsptwo}$ is around $\sim700$ GeV and $\sim 850$ GeV for $\mlspone\sim50$ GeV, in $Wh$ mediated $1\ell + (\geq 1b)+(\geq 1j)+2\gamma+\met$ and $WZ$ mediated $3\ell+(\geq 1b)+\met$ channels, respectively. Our results provide a roadmap of the currently available RPV parameter space that can be explored at the HL-LHC and demonstrate the impact of ML-based strategies in enhancing the sensitivity of electroweakino searches.

\subsection*{Acknowledgement}
  The authors would like to thank Najimuddin Khan for the fruitful discussions regarding the analysis. The work of RKB is supported by the World Premier International Research Center Initiative (WPI), MEXT, Japan, and by JSPS KAKENHI Grant Number JP24K22876. Some of the computation for this work was performed using resources at Kavli IPMU. A. Choudhury acknowledges Anusandhan National Research Foundation (ANRF) India for the Core Research Grant no. CRG/2023/008570. S. Sarkar acknowledges Anusandhan National Research Foundation (ANRF) India for the financial support through the Core Research Grant No. CRG/2023/008570.

\vspace{1cm}
\noindent
\textbf{Author contribution statement:} All authors have contributed equally.

\providecommand{\href}[2]{#2}\begingroup\raggedright\endgroup
\end{document}